%% file: dlp-arxiv.tex
\documentclass{ieeeaccessarxiv}

\if0
\makeatletter
\let\MYcaption\@makecaption
\makeatother
\usepackage{subcaption}
\makeatletter
\let\@makecaption\MYcaption
\makeatother
\fi

\usepackage{cite}
\usepackage{dcolumn}
\usepackage{bm}
\usepackage{amsmath,amssymb,amsfonts}
\usepackage{enumitem}
\usepackage{physics}
\usepackage{multirow}
\usepackage{graphicx}
\usepackage{blkarray}
\usepackage{xspace}
\usepackage{tabularx}
\usepackage[T1,T2A]{fontenc}
\usepackage{siunitx}
\usepackage{balance}
\usepackage[normalem]{ulem}
\usepackage{xspace}
\usepackage{color}
\def\BibTeX{{\rm B\kern-.05em{\sc i\kern-.025em b}\kern-.08em
    T\kern-.1667em\lower.7ex\hbox{E}\kern-.125emX}}

\usepackage{bmpsize}
\usepackage{algorithm}

\usepackage{amsmath}
\usepackage{amsfonts}
\usepackage{bm}
\usepackage{algorithm,algorithmic}
\usepackage{multirow}

\newtheorem{theorem}{Theorem}

\newtheorem{definition}{Definition}
\newtheorem{problem}{Problem}

\PassOptionsToPackage{hyphens}{url}
\usepackage{url}

\usepackage{qcircuit}
\usepackage{xspace}

\newcommand{\ggate}[1]{*+<1.4em>{\phantom{#1}} \POS ="i","i"+R; **\dir{-}; \qw}

\newcommand{\bvec}{\bm{b}}

\newcommand{\xvec}{\bm{x}}
\newcommand{\yvec}{\bm{y}}

\renewcommand{\ket}[1]{|#1 \rangle}

\newcommand{\circuitgen}{\textrm{CircuitGen}}
\newcommand{\postprocess}{\textrm{PostProcess}}
\newcommand{\device}{\textrm{Device}}
\newcommand{\ideal}{\textrm{Ideal}}
\newcommand{\dev}{\textrm{Dev}}
\newcommand{\noisy}{\textrm{Noisy}}
\newcommand{\unif}{\textrm{Unif}}
\newcommand{\kawasaki}{\textrm{Kawasaki}}
\newcommand{\ball}{\textrm{Ball}}

\renewcommand{\mod}{\textrm{mod}}
\renewcommand{\vol}{\textrm{vol}}
\newcommand{\covol}{\textrm{covol}}
\newcommand{\param}{\texttt{param}}
\newcommand{\rmspan}{\textrm{span}}

\newcommand{\mybeginproof}{\noindent{\bf Proof.} }
\newcommand{\myendproof}{$\Box$}

\renewcommand{\appendixname}{Appendix}

\RequirePackage{ifthen}

\newboolean{OmitPictures}   %Skip unnecessary pictures for the version to English proofreading service
\setboolean{OmitPictures}{false}

\newboolean{ShowComments}
\setboolean{ShowComments}{false}  % change to true to show remaining colorful author comments.
\ifthenelse{\boolean{ShowComments}}%
	{
		\newcommand{\ColorComment}[3]{%
				{\colorbox{#1}{\color{white}   \textsf{\textbf{#2}}} \textcolor{#1}{#3}}}%  Colorful box, initials, phrase
	}%
	{
		\newcommand{\ColorComment}[3]{}%  Do nothing at all
	}%

\definecolor{aonocolor}{rgb}{0,0,0.5}\newcommand{\aono}[1]{\ColorComment{aonocolor}{aono}{#1}}

\definecolor{rdvcolor}{rgb}{0,0.5,0}\newcommand{\rdv}[1]{\ColorComment{rdvcolor}{rdv}{#1}}
\definecolor{satohcolor}{rgb}{0.5,0,0.5}
\definecolor{michalcolor}{RGB}{255,127,80}
\definecolor{tanakacolor}{rgb}{1,0.6,0}
\definecolor{unocolor}{rgb}{0,0,1}

\usepackage{textcomp}

\begin{document}
\history{}
\doi{}

\title{The Present and Future of Discrete Logarithm Problems on Noisy Quantum Computers}

%\title{The Present and Future of DLP on Noisy Quantum Computers}

%\title{Analysing Discrete Logarithm Problems on Noisy Quantum Computers}

%%Discrete Logarithm Problems on NISQ computers by a quantitative way

%%A Quantitative measure
%%Post-processing algorithm
%%Success Probability

%Demonstrated
%Executing discrete logarithm problem on superconducting quantum processors

\author{
\uppercase{Yoshinori Aono}\authorrefmark{1},
\uppercase{Sitong Liu}\authorrefmark{2},
\uppercase{Tomoki Tanaka}\authorrefmark{3,5},
\uppercase{Shumpei Uno}\authorrefmark{4,5},
\uppercase{Rodney~ Van~ Meter}\authorrefmark{2,5} \IEEEmembership{Senior Member, IEEE},
\uppercase{Naoyuki Shinohara}\authorrefmark{1},
\uppercase{Ryo Nojima}\authorrefmark{1}
}
\address[1]{National Institute of Information and Communications Technology, 4-2-1 Nukui-kitamachi, Kogakei, Tokyo, 184-8795}
\address[2]{Faculty of Environment and Information Studies, Keio University, 5322 Endo, Fujisawa, 252-0882}
\address[3]{Mitsubishi UFJ Financial Group, Inc. and MUFG Bank, Ltd., 2-7-1 Marunouchi, Chiyoda-ku, Tokyo 100-8388}
\address[4]{Mizuho Research \& Technologies, Ltd., 2-3 Kanda-Nishikicho, Chiyoda-ku, Tokyo, 101-8443}
\address[5]{Quantum Computing Center, Keio University, 3-14-1 Hiyoshi, Kohoku-ku, Yokohama, Kanagawa, 223-8522}
%\tfootnote{This material is based upon work supported by }

\markboth
{Aono \headeretal: DLP}
{Aono \headeretal: DLP}

\corresp{Corresponding author: Yoshinori Aono (email: aono@nict.go.jp).}

\begin{abstract}
	The discrete logarithm problem (DLP) is the basis for several cryptographic primitives.
	Since Shor's work, it has been known that the DLP can be solved by combining a polynomial-size quantum circuit and a polynomial-time classical post-processing algorithm.
	Evaluating and predicting the instance size that quantum devices can solve is an emerging research topic.
	In this paper, we propose a quantitative measure based on the success probability of the post-processing algorithm to determine whether an experiment on a quantum device (or a classical simulator) succeeded.
	We also propose a procedure to modify bit strings observed from a Shor circuit to increase the success probability of a lattice-based post-processing algorithm.
    We report preliminary experiments conducted on IBM-Quantum quantum computers and near-future predictions based on noisy-device simulations.
    We conducted our experiments with the {\tt ibm\_kawasaki} device and discovered that the simplest circuit (7 qubits) from a 2-bit DLP instance achieves a sufficiently high success probability to proclaim the experiment successful.
    Experiments on another circuit from a slightly harder 2-bit DLP instance, on the other hand, did not succeed, and we determined that reducing the noise level by half is required to achieve a successful experiment.
    Finally, we give a near-term prediction based on 
    required noise levels to solve some selected small DLP and integer factoring instances.
\end{abstract}

\titlepgskip=-15pt
\maketitle

\input{intro}

\input{background}

\input{impl}

\input{exp}
\input{expnoisesim}

\input{discussion}

\bibliography{dlp-arxiv}
%\bibliography{refsaonojp}
\bibliographystyle{IEEEtran}

\appendix

\input{missing-lattice}

\input{appenddlpcircuits}

\input{appendshor}

\section*{Acknowledgement}
%\aono{Need to add something?}
This work was supported by MEXT Quantum Leap Flagship Program Grant Number JPMXS0118067285 and JPMXS0120319794. The results presented in this paper were obtained in part using an IBM Qunatum quantum computing system as a project of the Quantum Computing Center, Keio University. The views expressed are those of the authors and do not reflect the official policy or position of IBM or the IBM Quantum team.

\EOD
\end{document}

%% file: intro.tex
\section{Introduction}

Since Shor \cite{Shor97} proved that a reasonably large quantum circuit 
can solve both the integer factoring problem (IFP) and discrete logarithm problem (DLP) efficiently, many followers have been discussing its effect and implementability,
and have been attempting to reduce the attack's resource costs.

One of the emergent topics in the cryptographic area is to predict when the progress of quantum computers threatens modern cryptosystems.
Extrapolation from data points, such as regression \cite{SJ20}, is a common method for predicting time.
To put the points accurately, 
we need to reach an agreement on the following tasks.

(1) The threat will be realized by the progress of quantum computers.
Finding the relation between the performance and calendar years is needed. 
This relation depends highly on the published roadmaps from the companies \cite{IBMQVdouble,IBMQroadmap2020,IonQroadmap2020,Googleroadmap,Honeywellroadmap} which are continuously updated.
In this paper, our focus is not on this topic.

(2) Connecting a quantum computer's performance value 
and the ``difficulty'' of the hardest problem instance that the computer can solve.
To make the connection clear, it is necessary to quantify the concept of ``solve'' to use cryptographic applications.
%we have to fix the sense of ``solve'' in a quantitative way 
%to utilize it for the cryptographic applications.

\subsection{Summary of our Contribution}

\noindent
{\bf Defining successful experiments in a quantitative way}:
We propose our formalization of success probability 
and successful experiments in our framework including circuit generation, quantum device execution, and post-processing.
More precisely, after fixing a problem instance and a quantum circuit,
execution on a quantum device outputs a set of bit strings.
The post-processing algorithm then takes a set of bit strings as input and returns a set of candidate solutions to the problem instance.
The success probability is defined by the probability that 
the set of candidates contains the desired solution.
This success probability can be defined on
bit strings from the ideal device (a noiseless device for a quantum circuit simulated by a classical computer), a noisy device (real or simulated),
and a virtual device that outputs uniformly random bit strings.
Thus, by measuring where the quantum device is 
between the ideal and the uniform in the sense of success probability,
we can measure the performance of the device.
Also, we propose to claim the success of the experiment if
the device success probability is higher than the median of the ideal and the uniform.

\medskip\noindent
{\bf DLP experiments on an IBM Quantum device}:
We use the framework described above to furnish some data points for future prediction of the security of the DLP over a finite field against quantum computers.
We present results from experiments in which we used an IBM Quantum device to solve selected 2 bit instances realized by 7 and 8 qubit circuits.
More precisely, by using the {\tt ibm\_kawasaki} device of quantum volume (QV) $=32$,
we observed that the simplest DLP instance $2^z \equiv 1\ (\mod\ 3)$ with the smallest quantum circuit (Instance I in Table~\ref{tab:ourinstances}) outputs meaningful bit strings and success probability higher than the threshold that defines the successful experiment in our sense.

We also experimented with slightly more complicated circuits
(Instances II and III in Table~\ref{tab:ourinstances})
from the instance $2^z \equiv 2\ (\mod\ 3)$.
They generate 
bit strings with a low success probability.
To improve the success probability, 
we propose a simple algorithm for modifying bit strings from the devices.
As a result, we discovered that the device success probability of 
the instance III circuit is slightly below the threshold
and that reducing the noise by half is required to claim success using our noisy-device simulation.
Table~\ref{tab:ourinstances} shows the specifics of our DLP instances and quantum circuits.
Also, the smallest circuit used in the experiment is illustrated in Figure~\ref{fig:dlp1circuit}.

\medskip\noindent
{\bf Near-future prediction}:
By the simulation, we also predict how much noise needs to be reduced 
to solve larger instances IV and V which are from
the DLP instance $4^z\equiv\ 2\ \mod\ 7$ and $3^z\equiv 4\ \mod\ 7$.
We discovered that reducing the noise level to about $1/10$th is required
to solve IV from a device that solves III.
Table~\ref{tab:necessarynoiselevel} summarizes the results.
With the historical trend of reducing averaged CNOT gate errors by 1/2 every year \cite{IBMQheavyhex}, instances IV and V are expected to be solved within the next five years.
To solve a larger instance of DLP, it requires a more strong quantum device and additional techniques such as quantum error correction which will be developed in the future.

\subsection{Related Work}

\noindent
{\bf Discrete logarithm problem (DLP) and its applications}:
DLP over a finite field is a computational problem believed for a long time to be classically hard and is used as a theoretical foundation of digital signature schemes \cite{Sch89,NISTFIPS186-4}.
Many academic areas, including quantum computing and classical cryptography, 
are investigating how resilient the schemes are against quantum computers and 
when they become compromised.
We note that some cryptographic systems based on 
the DLP over elliptic curves (ECDLP) have also been used \cite{Mil86}.
Finding the smallest instance of ECDLP that is executable on quantum devices 
and experiment is a challenging open problem as of 2021.

\medskip\noindent
{\bf Experiments on Shor's algorithm on quantum devices}:
Many experiments that execute Shor's factoring circuits (and its subcircuits) have been performed \cite{LBC+12,MNM+16,ASK19} by using several quantum devices.
The latest record of factoring by a quantum circuit is $21=3\times 7$ by Amico et al. \cite{ASK19} by using the {\tt ibmqx5} device, and they also reported $35$ is infeasible.

However, there are no reports on DLP, even though the circuit  construction is very similar.
We discovered that the DLP is better suited for quantum benchmarking experiments in the NISQ era 
because many circuits that are simpler than factoring 15 or 21
can be constructed by changing three integers in DLP instances.
For instance, the numbers of CNOT gates are given in Table~\ref{tab:ourinstances} for DLP, and in Table~\ref{tab:ourshorinstances} for factoring.

Most of the existing experiments have employed folklore techniques, which are used in logical circuit optimization, to simplify the quantum circuits.
Because the latest quantum devices may not be able to execute modular arithmetic with sufficient accuracy \cite{OTU+19eng}, our circuit implementations are also simplified from the full implementation.
Following this simplifying policy, we carefully designed our circuit so that no information on the solution was used, as it was claimed that some experimental circuits were oversimplified by using problem solution information \cite{SSV13}.
Our circuits are made up of shift operations over qubits with no auxiliary bits.
Details are explained in Section~\ref{sec:simplegadgets}.

It is also necessary to follow the policy 
when we construct a post-processing algorithm
that generates a set of candidate solutions by using bit strings from quantum observations.
We also carefully designed our modification algorithm
that transforms an observed bit string to another bit string if it is necessary.

\medskip\noindent
{\bf Quantification of experimental results}:
It has been a long-standing problem 
to claim an experiment has succeeded or failed quantitatively.
Several quantifier functions have been used to estimate the quality of outputs from quantum devices.

Satoh et al. used the Kullback-Leibler divergence \cite{SOV20}.
Cross et al. \cite{CBS+19} used the probability that the outputs bit strings are in the heavy output set.
We found that these works share a spirit that compares 
distances between the ideal outputs, device outputs, and uniform random using their distance function; for more information, see Sections~\ref{sec:defpsucc} and 
\ref{sec:backgroundofourdef}.
We follow their approach and define our distance function based on the success probability of our generic framework including the required post-processing algorithm.

Another way of benchmark has been proposed.
The linear cross-entropy benchmark (linear XEB)
was used to claim demonstration of quantum supremacy by Arute et al. \cite{AAB+19}.
However, 
Barak et al. claimed that the linear XEB was fooled \cite{BCG20}.
The square of the statistical overlap is used in Amico et al. \cite{ASK19} 
which is introduced in Monz et al. \cite{MNM+16} to claim their advantage from previous works.

\subsection{Paper Organization}

In Section~\ref{sec:background}, we give a theoretical introduction to DLP, lattices,
and an overview of Shor's algorithm for solving DLP and a computational problem to recover the solution.
In Section~\ref{sec:framework}, we define our discussion framework, which including circuit generation, device execution, and post-processing.
In addition, with our motivation, we define a quantitative method for determining whether an experiment has succeeded or failed.
Section~\ref{sec:recoverdual} introduces modular-exponentiation gadgets used in our experiments, and lattice-based post-processing algorithm.
Appendix~\ref{app:backtheoryofPP} contains background theory on this post-processing algorithm.
Section~\ref{sec:experimentquantum} is the experimental section
that gives our experimental results on IBM Quantum.
Section~\ref{sec:prediction} gives simulation results of noisy quantum devices and comparison with the real device.
Section~\ref{sec:discussion} gives the concluding remarks and future work.

%% file: background.tex
\section{Background Theory}

\label{sec:background}

\rdv{A little clearer on dimension of Ball and the later lattice would be good.}
\aono{I'm changing $n$ to $K$ to represent the lattice dimension}
We introduce notations and background theory.

$\mathbb{N},\mathbb{Z},\mathbb{Q},\mathbb{R}$ are 
the set of natural numbers, integers, rational numbers, and real numbers, respectively.
For a prime number $p$, 
$\mathbb{Z}_p = \{ 0,\ldots,p-1\}$ is the field under modulo $p$.
$[n]$ denotes the set $\{ 1,2,\ldots,n \} $ for $n\in \mathbb{N}$.
$\ball_K(\bm{x},\rho)$ is the Euclidean-ball of radius $\rho>0$ with center $\bm{x} \in \mathbb{R}^K$ and $V_K(\rho)$ denotes its volume.

The Kullback-Leibler (KL) divergence
defined over two discrete probability distributions $P$ and $Q$ is 
\[
	D_{KL}(P||Q) := \sum_{i=1}^N P(x_i) \log \frac{P(x_i)}{Q(x_i)}
\]
where $P(x_i), Q(x_i)$ are the probability densities at $x_i$.
It is used to measure ``a distance'' from $P$ to $Q$ though
it is asymmetric and the triangle inequality does not hold .
We note that $D_{KL}(P||Q) = 0$ if and only if $P=Q$. 

\subsection{Discrete Logarithm Problem over a Field}

The DLP considered in this paper is the version defined over a prime field $\mathbb{Z}_p$ while Shor's algorithm can work over more general situations.
An instance of DLP is given by a tuple $(g,a,p) \in \mathbb{N}^3$
that represents the equation 
\begin{equation}
    \label{def:DLPinstance}
    g^z \equiv a\ (\mod\ p)
\end{equation}
to find $z\in \mathbb{Z}_p$.
Here, $g$ is assumed to be a generator under modulo $p$, that is, it satisfies
$g^{n}=1\ (\mod\ p)$ for $n=p-1$ and $\ne 1$ for any $n \in [p-2]$.

A variant of the DLP where one has extra information, e.g., the upper bound of $z$, has been considered under a cryptographic context and it significantly reduces the classical complexity to solve the problem \cite{Cheon10,EH17}.
In this paper we assume that no information is provided except for the DLP instance.

\subsection{Lattices}

We provide a brief overview of the lattices used in the analysis of Shor's algorithms on DLP.
Bremner's textbook \cite{Bre11} provides a gentle introduction.
The use of lattices in post-processing to solve 
the DLP by Shor's algorithm is also discussed in \cite{EH17,Eke18}.

For a sequence of (not necessary independent) vectors $\bm{b}_1,\ldots,\bm{b}_K \in \mathbb{Q}^m$, 
the lattice spanned by them is defined by the set 
\[
	L(B) := \left\{\sum_{i=1}^K a_i \bm{b}_i : \forall i, a_i \in \mathbb{Z} \right\},
\]
where $\bm{b}_1,\ldots,\bm{b}_K$ are called the basis vectors.
A vector in $L(B)$ is represented by row vectors.
We use the matrix $B :=[\bm{b}_1^T,\ldots,\bm{b}^T_K]^T$
to represent the basis in this paper.
Particularly, we say the lattice is full-rank if $K=m$ and the vectors are 
all independent.

A point $\bm{x} \in \mathbb{Q}^m$ is called a lattice point 
if $\bm{x} \in L(B)$.
Many useful lattice algorithms takes a matrix representation of
a lattice basis as input,
and the majority of them assume an input basis is a set of independent vectors,
though non-independent bases are occasionally used in applications.
There is an efficient algorithm for converting 
a non-independent basis to an independent basis that spans the same lattice
(See \cite[Sect.~6]{Bre11} or \cite[Sect.~2.6.4]{Coh13}). 
We assume the existence of such an algorithm in post-processing.

For a given basis, the fundamental region is defined by 
\[
    P(B) = 
    \left\{ \sum_{i=1}^K \alpha_i \bm{b}_i : \forall i, \alpha_i \in [0,1) \right\}
\]
and the covolume $\covol(L)$ of a lattice is defined by the volume of the region.
Both the lattice and its fundamental region are subsets of 
\[
    \rmspan(B) := \left\{ \sum_{i=1}^K \alpha_i \bm{b}_i : \forall i, \alpha_i \in \mathbb{R} \right\}.
\]
Note that $\covol(L)$ is efficiently computable from the basis 
if the basis vectors are independent,
since it is equal to $\displaystyle \prod_{i=1}^K \| \bm{b}^*_i \|$ where 
$\bm{b}^*_i$ is the vector in the Gram-Schmidt basis.

The Gaussian heuristic of lattices under the context of this paper 
claims that for an $n$-dimensional full-rank lattice $L$ and a ball $Y$,
the number of lattice points in $Y$ is expected by $\vol(Y) / \covol(L)$.
We should note that the original version of the Gaussian heuristic \cite{Ro56} argued for  a probabilistic distribution over random lattices, whereas the lattice discussed in the post-processing of Shor's algorithm may not be random.
Appendix~\ref{app:selectingradii} contains experimental evidence in small dimensions.

For a lattice $L$, its dual lattice $L^\times$ is defined by
the set 
\[
	L^\times := \{ \bm{x} \in \rmspan(L) : \forall \bm{v} \in L, \langle \bm{x},\bm{v} \rangle \in \mathbb{Z} \}
\]
where $\langle \bm{x},\bm{v} \rangle$ denotes the standard Euclidean inner-product.
A basis of $L^\times$ is explicitly given as $B(B^TB)^{-1}$ where $B$
is a basis of the primal lattice $L$.
It is simplified by $(B^T)^{-1}$ if $B$ is a square matrix.

For a given lattice basis $B$ in the $m$-dimensional space and a target vector $\bm{y} \in \mathbb{Q}^m$, the closest vector problem (CVP) is the computational problem to find a lattice point $\bm{x} \in L$ 
that minimizes $\| \bm{x} - \bm{y} \|$.
Anyone is allowed if there are many lattice points to minimize. 
a similar problem is the bounded distance decoding (BDD) problem.
For a problem instance $(B,\bm{y})$ and a bound $\rho > 0$, the goal is to find all lattice points $\bm{x} \in L$ such that $\| \bm{x}-\bm{y} \| \le \rho$.
If there is no vector in the ball $\ball_m(\bm{y},\rho)$, the empty symbol $\perp$ is returned.
We should point out that the version of BDD presented above differs significantly from the standard one.
In the typical situation of BDD, it assumes the existence of a unique solution $\bm{x}$.

\subsection{Overview of Shor's Algorithm in our Framework}

\label{sec:shor}

Fix a DLP instance (\ref{def:DLPinstance})
represented by $I=(g,a,p) \in \mathbb{N}^3$.
The parameter to specify the circuit size is $\param=(n_x,n_y)$.
$n_F$ is the bit size of the DLP instance, i.e., we set $n_F = \lceil \log_2 p \rceil$.
$n_x$ and $n_y$ are the circuit size parameters ($\ge n_F$) to define the size of 
Quantum Fourier Transform (QFT) and let $N_x := 2^{n_x}$ and $N_y := 2^{n_y}$.
Note that $n_x,n_y > 2n_F$ is necessary in theory, however, 
the computing device we used cannot operate the necessary number of qubits as of now.
Thus, we will use smaller numbers $n_x,n_y \approx n_F$ as in Table~\ref{tab:ourinstances}.

An overview of the quantum circuit for solving the DLP 
is shown in Figure~\ref{fig:dlpcircuit}.
We do not consider the serializing implementations \cite{HRS17,ASK19} that can reduce the number of qubits in the exponent part $n_x+n_y$ to one,
because our experimental environment (see, Section~\ref{sec:environment}) does not fully support such an implementation.

\Figure[h](topskip=0pt,botskip=0pt,midskip=0pt)[width=8cm]{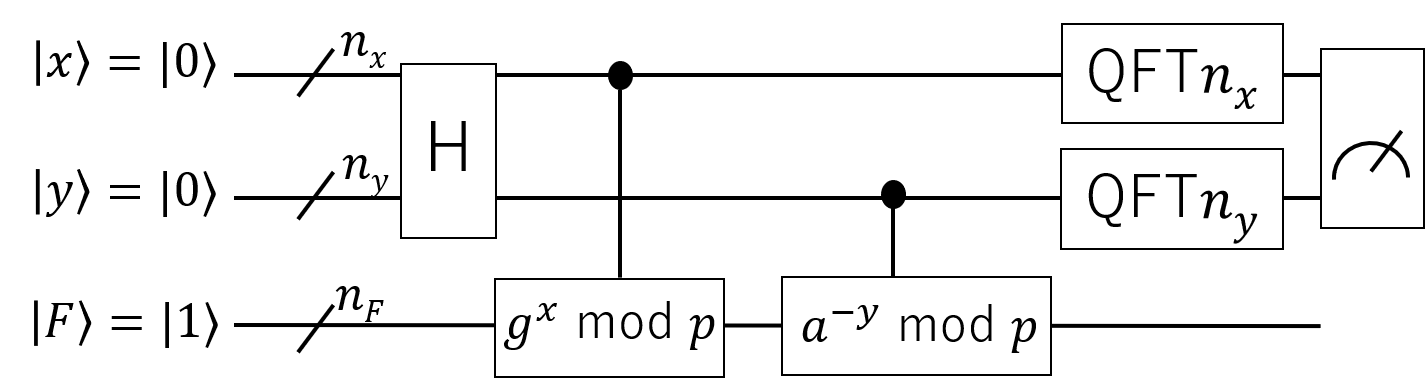}
    {Overview of Shor's circuit for solving DLP comprises Hadamard gates, two controlled modular exponentiations, and two quantum Fourier transforms
     \label{fig:dlpcircuit}}
% \begin{figure}[h]
% 	\centering
% 	\includegraphics[width=8cm]{dlpcircuit.png}
%     %\caption{caption test} %% <- Undefined control sequence??
%
% 	 \caption{Overview of Shor's circuit for solving DLP comprises Hadamard gates, two controlled modular exponentiations, and two quantum Fourier transforms \label{fig:dlpcircuit}}
% \end{figure}

It is easy to see that the bivariate function $F(x,y) := g^x a^{-y} \ \mod\ p$ 
has two periods $(p,0)$ and $(z,1)$ where $z$ is the desired DLP solution.
Thus, finding the periods will reveal the solution.
The circuit is designed to compute
the superposition of $\sum_{x,y} \ket{x,y,F(x,y)}$ over the box $[0,N_x-1] \times [0,N_y-1]$, and apply the QFT
for finding the periods.

The ideal state that assumes no quantum error, which 
corresponds to the distribution $P_{\ideal}$ in our framework,
is computed as follows.
The initial state $\ket{x,y}\ket{F} = \ket{0,0}\ket{1}$ is spread by the Hadamard gates.
\[
    \ket{0,0}\ket{1} \rightarrow 
    \sum_{x=0,y=0}^{N_x-1,N_y-1} \ket{x,y} \ket{1}
    \ \ \ (\textrm{ignoring nomalization}.)
\]

Then the modular exponent gates set the superposition
\[
	\rightarrow 
	\sum_{x=0,y=0}^{N_x-1,N_y-1} \ket{x,y} \ket{ g^x a^{-y}\ \mod\ p }.
\]
Finally, by applying the QFT circuits over $\ket{x}$ and $\ket{y}$,
the state to be measured is
\begin{equation}
    \label{eqn:stateafterqft}
\longrightarrow \sum_{x=0,y=0}^{N_x-1,N_y-1}  \sum_{k,\ell} e^{2\pi i \cdot \left( \frac{kx}{N_x} + \frac{\ell y}{N_y} \right)  }  \ket{k,\ell} \ket{ g^x a^{-y}\ \mod\ p }.
\end{equation}

An observed bit string $\bm{s}$ is 
represented as a pair $(k,\ell)$ of integers within the box $[0,0]\times [N_x-1,N_y-1]$.
It is also interpreted as the point in $[0,1)\times [0,1)$.
\begin{equation}
	\label{eqn:measurep}
	\bm{p} :=
	(p,r) = \left( \frac{k}{ N_x} ,\frac{\ell}{N_y} \right).
\end{equation}

\noindent
{\bf Computational Problem in Post-Processing}~
A post-processing algorithm tries to generate a set 
of candidates $z_1,\ldots,z_J$ of the solution to the instance $I$.
There could be several strategies for recovering.

Formally, the computational problem we have to consider after the quantum observation is as follows.
Details to derive the following problem will be provided in Appendix~\ref{app:backtheoryofPP}.

\begin{problem}
    \label{prob:duallatticeproblem}
    Let $L$ be the 2-dimensional integer lattice spanned by 
    $(p-1,0)$ and $(z,1)$ where $p$ is known and $z$ is unknown.
   $\left\{ \left(\frac{k_i}{N_x},\frac{\ell_i}{N_y} \right) \right\}_{i=1,\ldots,K}$ are the points from the observations,
    which are (noisy) approximations of lattice points in $L^{\times}$.
    Then, find (a suitable approximation of) $z$.
\end{problem}

We remark that it is enough to assume the existence of an algorithm 
that finds an approximation of $z$, say recover $z'$ s.t. $|z-z'| < z/4$
where the notation $|\cdot|$ is considered under modulo $p$.
With the approximated solution, consider the new DLP instance $(g,a\cdot g^{-z'},p)$ which has 
a smaller solution $z-z'$.
The approximation algorithm returns $z''$ s.t. $|z-z'-z''| < |z-z'|/4 < z/16$.
Repeating this process, we can recover the desired $z$.

An observed point is shifted from a lattice point by two factors.
The first is due to the finiteness of QFT, which is represented by a closed, but difficult to analyse formula precisely described by (\ref{eqn:probsumafterqft}) in Appendix~\ref{app:backtheoryofPP}. 
The other is caused by noise in quantum devices.
If the quantum circuit works without any quantum noise, 
we know the observed points are close to lattice points with a high probability \cite{Eke18,Eke19}.

\section{Generic Framework and Success Probability}

\label{sec:framework}

We revisit the standard procedures of computation by using quantum devices and propose our definition of the success of experiments and success probability.

\subsection{Three-Step Framework}

Our standard procedure to solve a computational problem using a quantum computer has the following three steps.

\begin{tabular}{lp{7cm}}
    (1) & Generate a quantum circuit from a given problem instance \\
    (2) & Execute the circuit on a quantum device \\
    (3) & Recover solution candidates and check.
\end{tabular}

In the NISQ era,
(1) and (3) are assumed to be classical (probabilistic polynomial) algorithms  which 
we may call pre-processing and post-processing, respectively.
If the performance of quantum devices improves, they can be replaced by quantum algorithms. 
However, we think considering this change is pointless for the time being, and 
we have left it as an open problem for the future.

Furthermore, we assume that the quantum circuit generated in (1) 
is optimized to solve the given problem instance rather than being designed to solve generic instances of a fixed size.
It has the potential to simplify the circuit significantly more than the circuit designed for generic instances.

More formally, we write the algorithms using the notations shown below.

\begin{tabular}{lp{7cm}}

$\bullet$& $\circuitgen(I,\param) \rightarrow QC$; A probabilistic algorithm that for given a problem instance $I$ and auxiliary information $\param$ such as the qubit size of the circuit,
it outputs a quantum circuit $QC$.
The output can be different each time. \\
$\bullet$&
$\device(QC) \rightarrow \bm{s}$; Execute the circuit $QC$ by a quantum device and get a bit string $\bm{s}$.
Here a device is considered to be either a real quantum device or a quantum simulator on a classical computer. \\
$\bullet$& $\postprocess(\bm{s}_1,\ldots,\bm{s}_K; I,\param) \rightarrow Z$;
A probabilistic algorithm that for given bit strings, problem instance $I$ and $\param$, it outputs a set of solution candidates $Z = \{ z_1,\ldots,z_J \}$ to $I$.
Here, the numbers $J,K$ of solution candidates and bit strings are not fixed, but assumed to be a polynomial of the instance size.
\end{tabular}

We have to add some remarks on the post-processing step.
The step could be further divided into two steps: 
modification of bit strings and recovering a solution.
The latter will be discussed as a lattice-based algorithm  in Section~\ref{sec:ourpostprocess}.
The former attempts to transform a bit string into a better one (a simple method will be presented in Section~\ref{sec:bitselection}) or to modify a probabilistic distribution.
One well-known technique for improving probability distributions is error mitigation 
\cite{temme2017error,endo2018practical}, which recovers a probabilistic distribution from an approximated distribution of bit strings derived from many shots and some information on the error distributions derived from additional experiments.
We do not consider the modification of the probabilistic distribution because our preliminary experiments using error mitigation do not significantly change the success probability.
Also, in the situation where we want to solve a large DLP instance, 
execution costs mean that it will be feasible to execute only a few shots
so that we cannot estimate the distribution.
Therefore, under the cryptographic contexts, we only consider 
the modification of bit strings.

\subsection{Defining Success Probabilities}

\label{sec:defpsucc}

\rdv{In all three of these subsections, III.A, B, C, it's only indirectly said that the QFT in Shor's algorithm gives a probabilistic amplitude distribution that means that there is a tradeoff between the amount of classical effort in post-processing and the number of shots. This needs to be stated a little more directly, probably along with some references.}
\aono{I added texts at the last of the above subsection.}

Within the framework in the above section, 
we can define the ``success'' for the experiments using real devices.
We start the discussion by introducing the devices that we will compare.

\begin{tabular}{p{8cm}}
(1) \ideal: this outputs bit strings from the noiseless quantum circuit, \\
(2) \dev: this is a real quantum device to be tested, and  \\
(3) \unif: this outputs random bit strings uniformly.
\end{tabular}

\begin{definition}
    \label{def:devicesuccessprobability}
    Let us fix a post-processing algorithm, in particular, 
    the number of input bit strings $K$ is fixed.
    For a device 
    $X \in \{ \ideal,\dev,\unif \} $,
    we denote by $P_X$ the probability distribution 
    over the bit strings $\bm{s}_1,\ldots,\bm{s}_K$
    from $K$ execution of the device $X$ with the circuit $QC_i = \circuitgen(I,\param)$.
    The suffix $i$ denotes the $i$-th execution
    where we can usually reuse the same circuit for each execution.
    The probability includes the circuit generator's random coins and superposition in the device.
    
    Also, the success probability 
    on the computational problem is defined by
    the probability that the set of output candidates contains the desired solution $I$:
    \begin{equation}
        \label{eqn:defpsucc}
        p_{X}
        := 
        \Pr[ z_{sol} \in \postprocess(\bm{s}_1,\ldots,\bm{s}_K; I,\param) ].
    \end{equation}
\end{definition}

Fixing the generator and post-processing algorithm with 
the number of input bit strings, the success probabilities
$p_{\ideal}, p_{\dev}$ and $p_{\unif}$ are fixed.
We should point out that this definition is intended for the situation in which
a solution candidate can be easily checked,
such as the situations where the IFP, DLP, and other NP problems.

If a program can be executed without any noise on a device,
the output distribution and success probability are
expected to be the same as the ideal.
Increasing the quantum noise of execution, 
$p_{\dev}$ drops.
Also, if the bit strings approach uniform noise, $p_{\dev}$ 
approaches to $p_{\unif}$.
Thus, we can expect 
\[
    p_{\unif} < p_{\dev} < p_{\ideal}
\]
and the scaled value 
\[
    s = \frac{p_{\dev} - p_{\unif}}{p_{\ideal} - p_{\unif}} 
    \in (0,1)
\]
can be used to measure the performance of devices.

Following the common strategy of existing works that we named ``the median principle'' (see also the next subsection), we say the device succeeded to solve 
the problem if $s>0.5$.
In other words,
we can say the device experiment to solve a problem instance is succeed if
\begin{equation}
    \label{eqn:ourcondition}
    p_{\dev} > \frac{p_{\ideal} + p_{\unif}  }{2}.
\end{equation}

We observe that the above definition of success, which we may call ``the  success of device experiments as a computing algorithm'' is somewhat disconnected from the cryptographic context in the real world.
In the cryptographic area, they can claim success if there exists a trial such that
$z_{sol} \in \postprocess(\bm{s}_1,\ldots,\bm{s}_K; I,\param)$ even though superpolynomial time was wasted.

For small instances considered in the NISQ era,
classical simulations to compute the accurate value of $p_{\ideal}$ and $p_{\unif}$ 
are possible and the median principle is a good criteria 
to check whether the device a performance is adequate.
For the era of large scale quantum devices, we may assume 
$p_{\unif}\approx 0$ since the number of candidates is polynomial despite the
search space is being exponential.
For DLP \cite[Section~4]{Eke19}, we can assume 
the success probability $p_{\ideal}\approx 1$ for sufficiently large instances.
The above assumptions deduce the threshold $0.5$.
It turns out some lazy version of the median principle for a large scale device is smoothly connected from the exact version.

\subsection{Background of the Definition}

\label{sec:backgroundofourdef}

We should add some explanations to the above definition (\ref{eqn:ourcondition}) of success.
Following preliminary experiments,
we needed to use some quantitative measures to determine whether or not a quantum device produces meaningful results.
However, when we simply applied the existing framework 
to Shor's DLP algorithm, we discovered two major issues.

The first one is that a typical condition for success on DLP computation should be defined over multiple vectors since the post-processing algorithm must take at least two bit strings from a device.
However, the existing criteria on success are defined by using a distribution over a single bit string.

The second issue is how we define our quantitative measure
that can be used to determine the limitations of state-of-the-art quantum devices and can be used to forecast future device progress.
We found that several previous criteria to evaluate experiments aim to compare ``a distance'' among device output distribution $P_{\dev}$, the ideal $P_{\ideal}$ and the uniform $P_{\unif}$ \cite{CBS+19,SOV20}.
Also, they employed the median principle 
that decides the success of the experiment by comparing 
$d(P_{\dev},P_{\ideal})$ and $d(P_{\dev},P_{\unif})$
by using a reasonable distance function $d(\cdot,\cdot)$.

\begin{definition}
    (The median principle)
    Fixing the quantum circuit and some function $d$ 
    to measure the distance between two probability distributions.
    We say a device experiment is succeed if
    $d(P_{\dev},P_{\ideal}) < d(P_{\dev},P_{\unif})$.
\end{definition}

Therefore, we first define the success probability $p_{\device}$
based on the outputs from the post-processing algorithm,
and then define the successful experiment with the distance function 
$d(P,Q) = |p-q|$.
Here, $p$ and $q$ are the corresponding success probability of 
$P$ and $Q$, respectively.

We think our criteria as of now is a basic version 
and expect that followers will update it according to their own needs.

\subsection{Success Probability and KL-divergence}

\label{sec:psuccbackground}

\begin{figure}[htbp]
	\centering
\ifthenelse{\boolean{OmitPictures}}{OMITPicture}{
    \includegraphics[width=4cm]{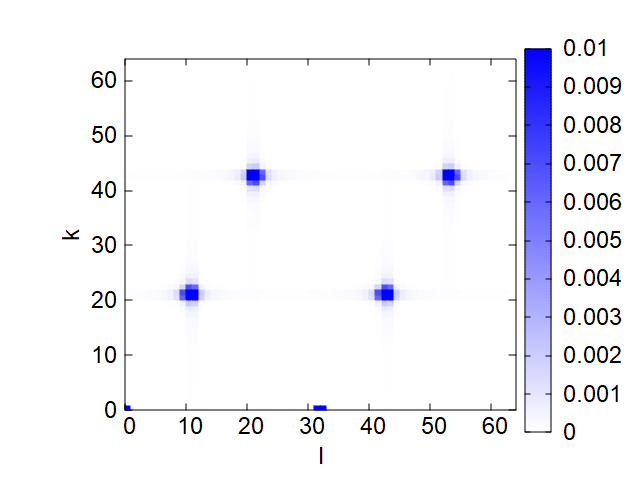}
	\includegraphics[width=4cm]{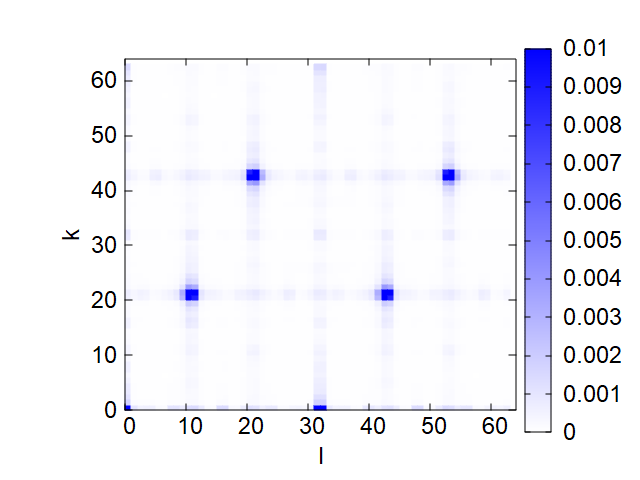}
}
	\caption{Density function from the exact simulation $P_{\ideal}$ and noisy simulation $P_{\noisy}$ with parameters $p_1=0.0003$ and $p_2=0.003$ that outputs similar data.
    Although we might think the device works well
    on first impression, the asymmetricity of the KL-divergence 
    drives us questionable results.
	\label{fig:twosimulations}}
\end{figure}

We give an experimental motivation to use the success probability 
as the function $d(P,Q)$ rather than the KL-divergence.
We simulated the probability density functions
of ideal and noisy execution for the quantum circuit instance VI (see Table~\ref{tab:ourinstances} in the experimental section).
In this case, we set the noise parameters to have a two-bit gate depolarizing error $p_2=0.003$
and a single-bit gate depolarizing error $p_1 = 0.1 \cdot p_2$.
Figure~\ref{fig:twosimulations} depicts a comparison of the ideal distribution and the noisy distribution.
The horizontal and vertical axes represent the position of $(\ell,k)$ as determined by 12 bit observations.

Although we might think the device works well
on first impression, 
the following KL-divergence metrics show it is questionable.
The concrete values are 
$D_{KL}(P_{\ideal},P_{\noisy}) = 1.022$,
$D_{KL}(P_{\noisy},P_{\unif}) = 2.881$,
by which we can conclude that device output is good.
On the other hand, the reverse directions are
$D_{KL}(P_{\unif},P_{\noisy}) = 1.745$,
$D_{KL}(P_{\noisy},P_{\ideal}) = 3.305$,
which are evidence that the device output is closer to the uniform than the ideal distribution.
Hence, we can obtain contradicting results.

%Our old comparison method: $D(\device || Ideal)$ vs. $D(Unif || Ideal)$

%p2=0.003
%KLdiv(Exact,Unif)=7.09098
%KLdiv(Unif,Exact)=7.05203

%KLdiv(Exact,E300)=1.02242
%KLdiv(E300,Unif)=2.88055
%KLdiv(Unif,E300)=1.74536
%KLdiv(E300,Exact)=3.30542

%p2=0.005
%KLdiv(Exact,Unif)=7.09098
%KLdiv(Unif,Exact)=7.05203

%KLdiv(Exact,E200)=1.7519
%KLdiv(E200,Unif)=1.71529
%KLdiv(Unif,E200)=1.078
%KLdiv(E200,Exact)=4.62675

From the result, we think a naive comparison of 
KL-divergence is not suitable to decide the success of experiments.
Furthermore, our goal is not only to benchmark quantum devices 
but also to measure the device performance as an accelerator for solving DLP.
This is why we proposed using 
the overall success probability to determine whether 
an experiment is succeed, as in (\ref{eqn:ourcondition}).

%% file: impl.tex
\section{Our Implementation}

In Section~\ref{sec:shor}, we omitted details of 
modular-exponentiation arithmetic gadgets and 
a post-processing algorithm to generate solution candidates to Problem~\ref{prob:duallatticeproblem}.
Many implementations have been proposed for their building blocks.
In this section, we give details of our version
of the implementation used in our experiments.

\label{sec:recoverdual}

\subsection{Our Quantum Modular-Exponentiation Circuit}

\label{sec:simplegadgets}

The modular exponentiation gadgets are the most complicated part of Shor's circuit.
They are typically implemented by a sequence of modular multiplication operations.

Based on our understanding of the latest hardware, the general modular multiplication circuit, even for problems involving only a few qubits, is too demanding for the latest generation of quantum machines \cite{OTU+19eng}.
As a result, we must consider specific cases 
and simplify the modular multiplication circuit as much as possible
by following existing reports.

We used the standard binary representation of nonnegative integers. 
For a state $\ket{F_{n-1}\cdots F_0}$ comprising $n$ qubits,
we regard it as the integer $F = F_{n-1}2^{n-1} + \cdots + F_0 $ and denote it as $\ket{F}$.
To simplify the circuits, we exploited the fact that the shift rotation 
$\ket{F_{n-1}\cdots F_0} \rightarrow \ket{F_{n-2}\cdots F_0F_{n-1}}$
computes the function $F \mapsto 2F\ \mod\ (2^n-1)$ \cite{MNM+16}
without using any auxiliary qubits.
A stack of the shift rotation circuits realizes the modulus power function $F \mapsto  F\cdot 2^k\ \mod\ (2^n-1)$
for any $k$.

\begin{figure}[htbp]
	\centering
	\includegraphics[width=8cm]{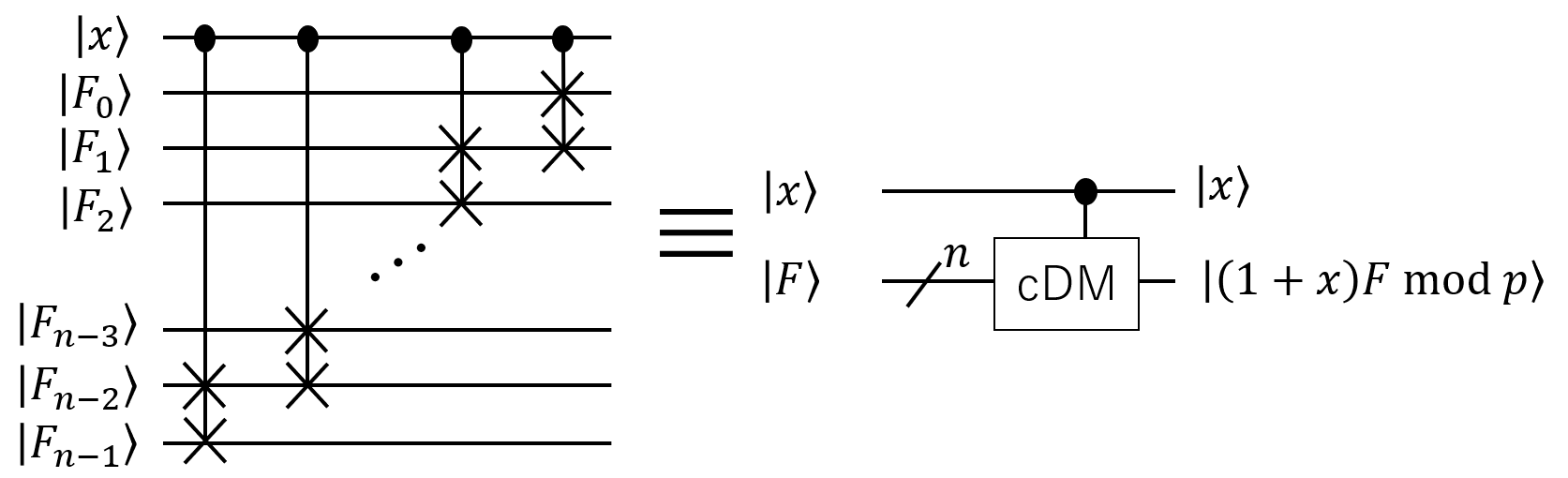}
	\caption{ $n$-bit double-then-modular quantum circuit 
		without any auxiliary qubits \label{fig:doublethenmod}}
\end{figure}

With this gadget, 
the controlled version of double-then-modular operation is
expressed as 
\[
    \begin{array}{ll}
    \ket{x}\ket{F} &
    \rightarrow \ket{x}\ket{(1+x)F\ \mod\ (2^n-1)} \\
    & \hspace{6mm} = 
    \left\{ 
    \begin{array}{ll}
    	\ket{F_{n-1}\cdots F_0}	& (x=0) \\
    	\ket{ F_{n-2}\cdots F_0 F_{n-1} } & (x=1) 	
    \end{array}
    \right.
    \end{array}
\]
whose circuit is shown in Figure~\ref{fig:doublethenmod}.

In the modular exponentiation, we considered 
quantum circuits that use only NOT, CNOT gates, and the above gadget.
In particular, such simplified circuits meet the following requirements.
(1) $p$ is a number of the form $2^n-1$ ($p$ should be prime in a cryptographic context), and 
(2) both $g^{2^j}\ \mod\ p$ for $j\ge 1$ and $(a^{-1})^{2^k}\ \mod\ p$ 
for $k\ge 0$ are represented by a power of two.

Thus, the modulus exponentiation part that computes $\ket{x,y,1} \rightarrow \ket{x,y,g^x a^{-y}\ \mod\ p}$ comprises of $\ket{x,y,1} \rightarrow \ket{x,y,g\cdot x_0}$ (implemented by NOT and CNOT gates, where $x_0$ is the least significant bit of $x$),
and the controlled double-then-modular gadgets.

An illustrative example is for instance $3^z \equiv 4\ (\mod\ 7)$
with $n_x=n_y=4$ (instance V in Table~\ref{tab:ourinstances}).
By the relations $3^2 \equiv 3^8 \equiv 2 \ (\mod\ 7)$,
$3^4 \equiv 4 \ (\mod\ 7)$, $4^{-1} \equiv 4^{-4} \equiv 2\ (\mod\ 7)$ and 
$4^{-2} \equiv 4^{-8} \equiv 4 \ (\mod\ 7)$ 
it derives the following calculation:
\[
	g^x a^{-y}\ \mod\ p = 3^{x_0} \cdot 2^{x_1} \cdot 4^{x_2} \cdot 2^{x_3} \cdot 2^{y_0}
	\cdot 4^{y_1} \cdot 2^{y_2} \cdot 4^{y_3} \ \mod\ 7.
\]
Thus, the computing circuit is started by $\ket{x,1} \rightarrow \ket{x,3^{x_0}}$ which is implemented by two CNOT gates,
and the other parts are double-then-modular gadgets.

\subsection{Our Post Processing}

\label{sec:ourpostprocess}

We present our version of a classical algorithm for recovering candidates of $z$ via dual lattices,
whose naive extension would be useful in many situations involving Shor-type algorithms.
See, for example \cite{EH17,Eke17,Eke18} for details on the algorithm that employs primal lattices.
In theoretical analysis, we assume, as in many previous works, 
that there is no gate or measurement noise during algorithm execution.

We outline the algorithm in Algorithm~\ref{alg:postprocess}.
For the input of a sequence of bit strings $\bm{s}_1,\ldots,\bm{s}_K$,
it outputs a set of solution candidates $z_1,\ldots,z_J$
to the DLP instance.
As we described in Appendix~\ref{app:backtheoryofPP}, 
the bit strings correspond to 
the points $\{ (p_i,r_i) \}_{i=1,\ldots,K}$ are approximations of
points in $L^\times \cap [0,1)^2$ where $L^\times$
is the dual lattice of $L$ spanned by $(p-1,0)$ and $(z,1)$
with the DLP solution $z$.

Also, the lattice defined by the $(K+1)\times K$ matrix
\begin{equation}
	\label{eqn:defB}
    \begin{small}
    B :=
    \left[
    \begin{array}{c}
        \bm{b}_1 \\
        \bm{b}_2 \\
        \bm{b}_3 \\
            \vdots \\
        \bm{b}_{K+1}
    \end{array}
    \right] 
    =
    \left[
    \begin{array}{ccccc}
    	p_{1} & p_{2} & \cdots & p_{K} \\
    	1 & 0 & \cdots & 0 \\ 
    	0 & 1 & \cdots & 0 \\
    	\vdots & \vdots & \ddots & \vdots \\ 
    	0 & 0 & \cdots & 1 
    \end{array}
    \right],
    \end{small}
\end{equation}
has a lattice point $\bm{x} = \sum_{i=1}^{K+1}a_i \bm{b}_i$ close to the target vector 
\begin{equation}
    \label{eqn:defy}
    \yvec = (r_{1},r_{2},\ldots,r_{K}).
\end{equation}
Finding $\bm{x}$, we can recover the DLP candidate solution $z$
since the combination coefficient $a_1$ is $-z$.
The details of the post-processing are given in Appendix~\ref{app:backtheoryofPP}.

Therefore,
for a given point set $\{ (p_i,r_i) \}_{i=1,\ldots,K}$,
executing a BDD subroutine $BDD(B,\bm{y},\rho_K)$ generates 
a list of vectors $\bm{v}_1,\ldots,\bm{v}_J$
and the corresponding combination coefficients derive
the candidate set $z_1,\ldots,z_J$.
The parameters in the BDD subroutine are $B$ and $\bm{y}$, which are defined above.
The remainder of this section is concerned with the selection of the searching radius $\rho_K$.

We need to set it so that 
the overall success probability is sufficiently high
while keeping the computing time feasible.
In other words, 
we need to keep the number of lattice points in $\ball_K(\yvec,\rho_K)$ small, namely, $O(K)$.

To bound the number, the Gaussian heuristic assumption 
provides us a good estimation in general.
The assumption claims that
the number of lattice points in $\ball_K(\yvec,\rho_K)$ 
is averageabout one if we set
\begin{equation}
    \label{eqn:setrho}
    \rho_K
    =V_{K}(1)^{\frac{1}{K}} \covol(B)^{\frac{1}{K}}
    = 2^{-n_x / K}
    \Gamma(K/2+1)^{1/K} \cdot \pi^{-\frac{1}{2}},
\end{equation}
for detail of the derivation, see Appendix~\ref{app:selectingradii}.
Although we basically followed Eker\aa's theory \cite{Eke19}
to implement our post-processing, 
we have found two issues to modify whose 
details are postponed to Appendix~\ref{app:latticeanalysis} and \ref{app:selectingradii}. 

The first one is the determinant computation of the lattice $B$.
In the lattice application, a lattice basis is usually given by a triangle matrix for the simplicity of determinant analysis.
In our situation, the lattice is represented by a $(K+1)\times K$ matrix as in (\ref{eqn:defB}).
An auxiliary column $\bm{c} = (\tau,0,\ldots,0)^T$ 
on the left of $B$, is commonly used to analyze such bases, 
where $\tau$ is a parameter to be optimized.
Because the dimension has been changed, it may cause some theoretical issues.
In Theorem~\ref{thm:covolB} of Appendix~\ref{app:latticeanalysis}, we provide a theoretical analysis of $\covol(B)$.
This is appropriate for our situation, in which experiments to solve the DLP are carried out using small dimensional lattices.

The second one is on the number of lattice points 
within $\ball(\bm{y},\rho_K)$.
We experimentally find that an exponential number of lattice points is contained in the ball (Appendix~\ref{app:selectingradii}). 
This result shows that the DLP lattices are not close to random enough required to use the Gaussian heuristic.
Thus, we should work in low dimensions such as $K+1 \le 10$ to limit 
the number of found vectors.
Unfortunately, to the best of our knowledge, how a method for setting  
the radius in a large dimension is unknown.

In small dimensions, 
the number of lattice points within the ball 
has a non-negligible variance.
We experimentally discovered that there is a non-negligible number of trials in which there is no lattice point within $\ball_K(\bm{y},\rho_K)$.
In this case, we add the execution of the CVP oracle that finds 
the closest lattice point to $\bm{y}$ and recovers a solution candidate from it.
As a result, Algorithm~\ref{alg:postprocess} describes the post-processing algorithm.
We can easily check the candidates by computing $g^{z_i}\ (\mod\ p)$.
We say the success of the experiment when one of the candidates passes the check.
If otherwise, the experiment is failed and we try with a new set of inputs.

\begin{algorithm}[h]
	\begin{small}
		\caption{Our Post-processing algorithm for DLP \label{alg:postprocess} }
		\begin{algorithmic}[1]
			\REQUIRE $\bm{s}_1,\ldots,\bm{s}_K$: non-zero bit strings from a device.
			\ENSURE  $Z = \{ z_1,\ldots,z_J \} $: a set of solution candidates for the DLP instance
            \STATE Convert $\bm{s}_i$ to $(p_{i},r_{i})$ for $i=1,\ldots, K$
            \STATE Construct the BDD instance $(B,\bm{y},\rho_K)$ in 
            (\ref{eqn:defB}), (\ref{eqn:defy}) and (\ref{eqn:setrho})
            \STATE $BDD(B,\bm{y},\rho_K) \rightarrow V = \{ \bm{v}_1,\ldots,\bm{v}_J \}$ \\
            \IF{$V=\phi$}
                \STATE $CVP(B,\bm{y}) \rightarrow V = \{ \bm{v}_1 \}$ 
            \ENDIF
            \STATE Recover solution candidates $z_i$ from $\bm{v}_i$ for $i=1,\ldots,J$
            \RETURN $z_1,\ldots,z_J$ 
		\end{algorithmic}
	\end{small}
\end{algorithm}

\subsection{Selection and Modification of bit strings}

\label{sec:bitselection}

In our preliminary experiments, 
we found a naive execution of Algorithm~\ref{alg:postprocess}
sometimes fails to find the desired DLP solution.
To increase the success probability, 
we propose two methods to select and modify the bit strings from 
the devices.

The first method is to remove the zeros.
It is based on the fact that the zero vector $\bm{p}_i = (0,0)$, 
derived from the zero bit string, 
is useless because the zero vector is always the point in the dual lattice $L^\times$.
As a result it provides no information
and we remove vectors representing points near $(0,0)$
from the post-processing algorithm's inputs.
Because the number of instances considered in our experiments is very small,
we removed only zero vectors whereas vectors near the zero should be removed in the case of large DLP instances.

The second method is using the properties of distributions after QFT.
Since $L^\times$ is spanned by two vectors in the matrix $D$ (see Appendix~\ref{app:backtheoryofPP} for details),
any points in $L^\times \cap [0,1)^2$ from any DLP instance considered under modulo $p$ must be in the set
\begin{equation}
    \label{eqn:defsp}
    S_p =
    \left\{ (0,0)\right\}
    \cup
    \left\{ \frac{1}{p-1}(c_1,c_2) : 
        \begin{array}{l} c_1 =1,\ldots,p-2 \\
            c_2=0,\ldots,p-2 
        \end{array}
        \right\}.
\end{equation}

Thus, if the converted point $\bm{p}'=(p',r')$ is not very close to 
a point in $S_p$, it must have bit flip errors.

We emphasize that determining whether a vector is close to a point in $S_p$ can be done quickly and without using any DLP solution information. 
Furthermore, unlike error mitigation techniques, it does not make use of any information from the probability distribution.

%The relation to the correct point $\bm{p}=(p,r)$ is
%
%\[
%    p' = p \pm 2^{i_1} \pm 2^{i_2} \pm \cdots \ \ \ \textrm{and}\ 
%    r' = r \pm 2^{j_1} \pm 2^{j_2} \pm 
%\]
For a bit string containing errors, we can try all the 1 bit flips to ensure that the corresponding vectors are on the correct points.
If all of the trials fail, we reject the bit string, and try 
all the 2-bit, 3-bit, and so on flips if necessary.
In our experiments, we try every single 1 bit flip.

Therefore, we modify the bit strings from devices as in Algorithm~\ref{alg:modifybits}.
Note that we use the criteria $\bm{p}\in S_p$ and $\bm{p}_i \in S_p$ exactly
to check the conditions in Step~\ref{step:if1} and \ref{step:if2}, respectively in our experiments. 

\begin{algorithm}[h]
	\begin{small}
		\caption{Our bit string modification algorithm for DLP \label{alg:modifybits} }
		\begin{algorithmic}[1]
			\REQUIRE $\bm{s}$: a bit string from a device
			\ENSURE $(p',r')$: modified vector
            \STATE Convert $\bm{s}$ to $\bm{p} = (p,r)$ as in (\ref{eqn:measurep})
            \IF{$\bm{p}$ is sufficiently close to a point in $S_p$ in (\ref{eqn:defsp})}  \label{step:if1}
            \RETURN $\bm{p}$ 
            \ENDIF
            \FORALL{$\bm{p}_i = (p',r')$ is from $i$-th bit flip of $\bm{s}$}
                \IF{$\bm{p}_i$ is sufficiently close to a point in $S_p$}   \label{step:if2}
                \STATE $C\leftarrow C \cup \{ \bm{p}_i \}$ 
                \ENDIF
            \ENDFOR
            \IF{$C=\phi$}
                \RETURN $\perp$
            \ELSE
                \STATE Choose randomly $\bm{p}$ from $C$
                \RETURN $\bm{p}$
            \ENDIF
		\end{algorithmic}
	\end{small}
\end{algorithm}

Using the above two modification methods in Step~1 in Algorithm~\ref{alg:postprocess}, each input bit string
is translated to a modified point $(p_i,r_i)$, or rejected sample $\perp$.
We replace $\perp$ with a new modified sample from a device and
move to Step~2 after all the points are legitimated.

%% file: exp.tex
\section{Experiments in IBM Quantum}

\label{sec:experimentquantum}

This section provides experimental results on a real superconducting quantum computer.
For Shor's quantum circuits to solve selected instances of DLP, 
we measured probability distributions and success probabilities.

\subsection{Experimental Environment}

\label{sec:environment}

\begin{figure}[htbp]
	\centering
	\includegraphics[width=8cm]{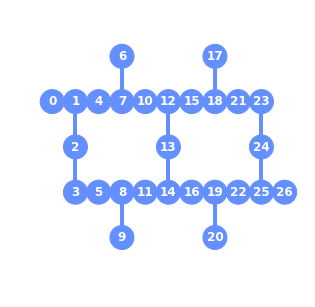} 
	\caption{connectivity of {\tt ibm\_kawasaki} }
	\label{fig:kawasaki}
\end{figure}

First, we explain the target DLP instances and the conditions of the experiments.
The experiments were performed with IBM Quantum device {\tt ibm\_kawasaki} which has the connectivity of qubits shown in Figure~\ref{fig:kawasaki}.\par

%To evaluate the performance of different size Shor's circuits to solve DLP instances, we embedded the selected 2 and 3-bit instances (Table~\ref{tab:ourinstances}).
Table \ref{tab:ourinstances} shows the DLP instances with the gadgets described in Section~\ref{sec:simplegadgets}.
$n_x$ and $n_y$ are the widths of the exponent variables, which are equal to 
the size of QFT gadgets.
$n_F$ is the number of qubits to compute $F(x,y)$.
Since we do not use any auxiliary bits to perform the computation under modulo $p$,
it consumes $\lceil \log_2 p \rceil$ bits.
Therefore, the total number of qubits used in the curcuit is $\# Q = n_x+n_y+n_F$.

%The parameters $n_x$ and $n_y$ are the width of the exponent variables, and $\#Q$ is the number %of qubits calculated by $n_x+n_y+n_F$.\tanaka{memo:nanka eigo ga hen}

\begin{table}[!htbp]
\caption{Summary of parameters for target DLP instances.
\label{tab:ourinstances}}

\begin{center}
	\begin{tabular}{|c|c|c||r|r||r|r|}
		\hline
		& DLP & DLP  & \multirow{2}{*}{$n_x$} & \multirow{2}{*}{$n_y$} & \multirow{2}{*}{$\#$Q}& \multirow{2}{*}{$\#${\tt cx}} \\
		& instance & bits & &&& \\
		\hline
		\hline
		I & $2^z=1\ \mod\ 3$ &\multirow{3}{*}{2} & 3 & 2 & 7 & 15   \\
		\cline{1-2} 	\cline{4-7} 
		II & \multirow{2}{*}{$2^z=2\ \mod\ 3$} & & 3 & 2 & 7 & 32 \\
		\cline{1-1} 	\cline{4-7} %cline{1} doesn't work 
		III &  && 3 & 3 & 8 & 38 \\
		\hline
		\hline
		IV & $4^z=2\ \mod\ 7$& \multirow{3}{*}{3} & 3 & 3 & 9 & 179 \\
		\cline{1-2} 	\cline{4-7} 
		V & \multirow{2}{*}{ $3^z=4\ \mod\ 7$} &  & 4 & 4 & 11 & 255 \\
		\cline{1-1} 	\cline{4-7} 
		VI & &  & 6 & 6 & 15 & * \\
		\hline
	\end{tabular}
\end{center}
\end{table}

\begin{figure}[htbp]
	\centering
	\includegraphics[width=8cm]{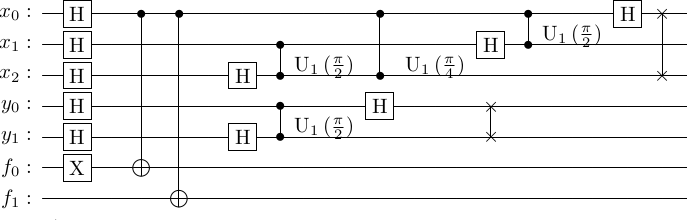} 
	\caption{Our abstract circuit of instance I}
	\label{fig:dlp1circuit}
\end{figure}

\begin{figure}[htbp]
	\centering
	\includegraphics[width=8cm]{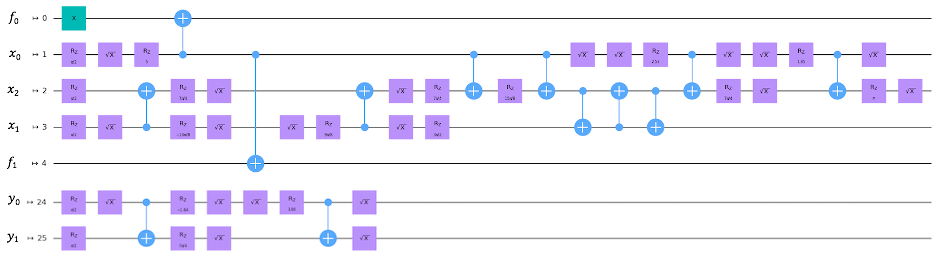} 
	\caption{Transpiled circuit of instance I after optimization by the {\tt transpile} command with optimization level 3. }
	\label{fig:dlp1circuittranspiled}
\end{figure}

The circuit of instance I assuming full-connectivity is shown in Figure~\ref{fig:dlp1circuit} as an illustrative example.
On the other hand, current superconducting quantum computers do not have fully connectivity.
Thus, we have to embed the circuit by fitting the topology of the target quantum device so that the performance loss is small as possible as we can.
There are many methods to measure the loss.
We tried to reduce the total number of CNOT ({\tt cx}) gates, which is expected to be equal to increase the total performance because the fidelity of CNOT gate is lower than single gate in IBM Quantum devices.
Such circuit-to-circuit translation is performed by the ${\tt transpile}$ command in Qiskit\cite{Qiskit}.
This command returns a variety of feasible circuits since it uses a random number in its optimizing subroutine.
We set the seed of random number generator as the input of the ${\tt transpile}$ command, 
so that we can generate many circuits and take the circuit whose number of CNOT gates is minimum.
We also set the options so that the output circuit consists of the gate set {\tt ['cx','id','rz','sx','x']}, and optimization level 3. 
Figure~\ref{fig:dlp1circuittranspiled} shows the circuit of instance I used in our experiments.

The circuits of instances II and III, which are used in our experiments, are shown in Figure~\ref{fig:dlp23circuit} in Appendix~\ref{app:dlpcircuit}.
As we explained in the next section, experiments using the both instances are not succeed because its success probabilities do not meet the level of threshold values.
Instances IV to VI are not executed because they are clearly more complicated than II and III,
and the success probabilities are very lower than the thresholds.

The summary of the experimental environment is shown in Table~\ref{tab:experimentsummary}.
One execution consists of $8,192$ shots and measurements,
and one experiment consists of 100 repeats of execution to evaluate the statistical variance.
Totally, we have $819,200$ bit strings for each instance.

%Denoted that, the number of shots (measurement) of the expriments is $8192$ and the same %experiments are repeated $100$ times to evaluate the statistical variance.
%Then, we also used the \{1, 3, 2, 24, 25, 0, 4 \}, \{1, 0, 4, 5, 8, 2, 3\}, and \{ 1, 0, 4, 5, 8, %11, 2, 3\} in instances I, II, and III, respectively.

\begin{table}[!htbp]
\caption{Summary of experimental environment and conditions.
\label{tab:experimentsummary}}

\begin{center}
	\begin{tabular}{|l|l|l|}
		\hline
            & qubits indexes & Date of experiment \\
		\hline
        I   & $\{1, 3, 2, 24, 25, 0, 4 \}$ & Jul 21th, 2021 \\ 
		\hline
        II   & $\{1, 0, 4, 5, 8, 2, 3\}$ &Jul 21th, 2021\\ 
		\hline
        III   & $\{ 1, 0, 4, 5, 8, 11, 2, 3\}$ &Jul 21th, 2021\\
        \hline
	\end{tabular}
\end{center}
\end{table}

\subsection{Results from real quantum devices}

\begin{figure}[htbp]
	\centering
	\includegraphics[width=8cm]{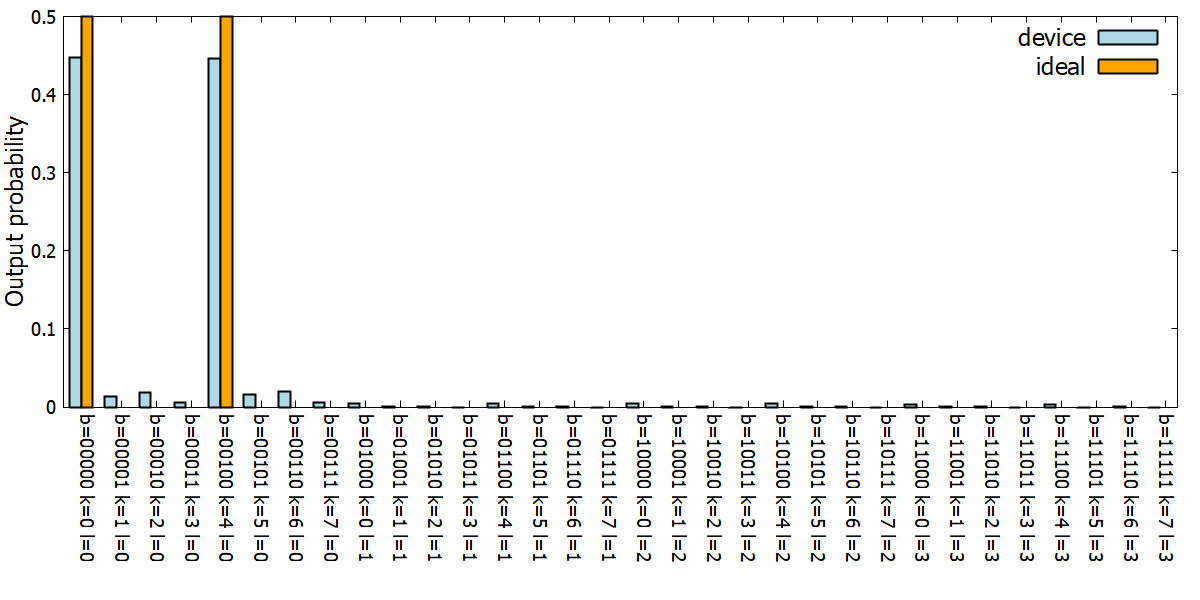} %instance I
	\includegraphics[width=8cm]{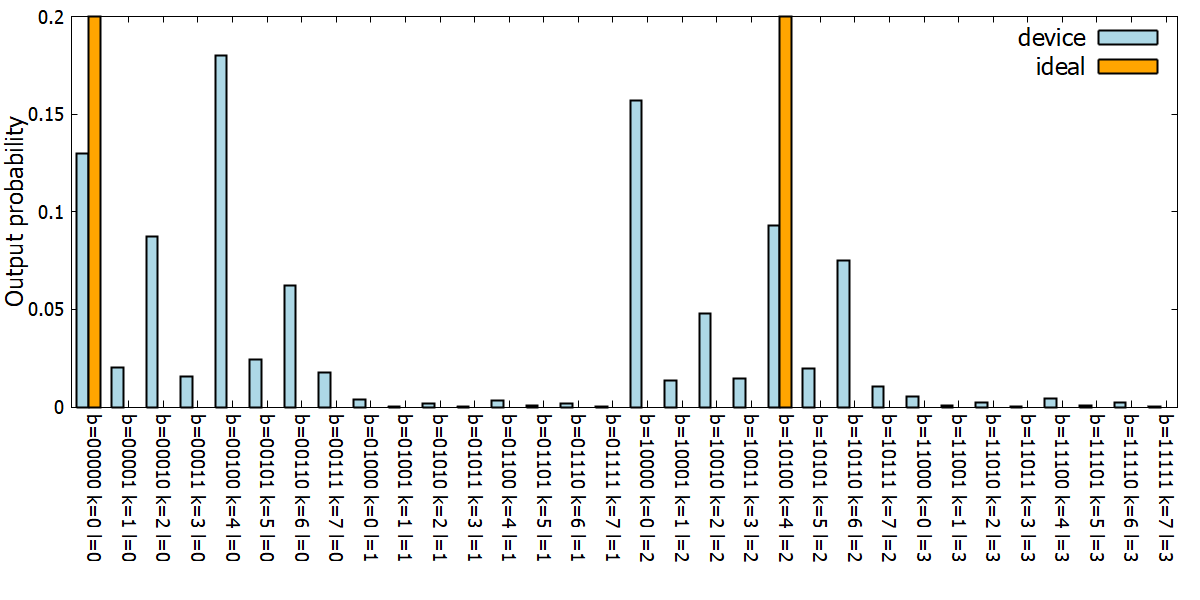} %instance II
	\includegraphics[width=8cm]{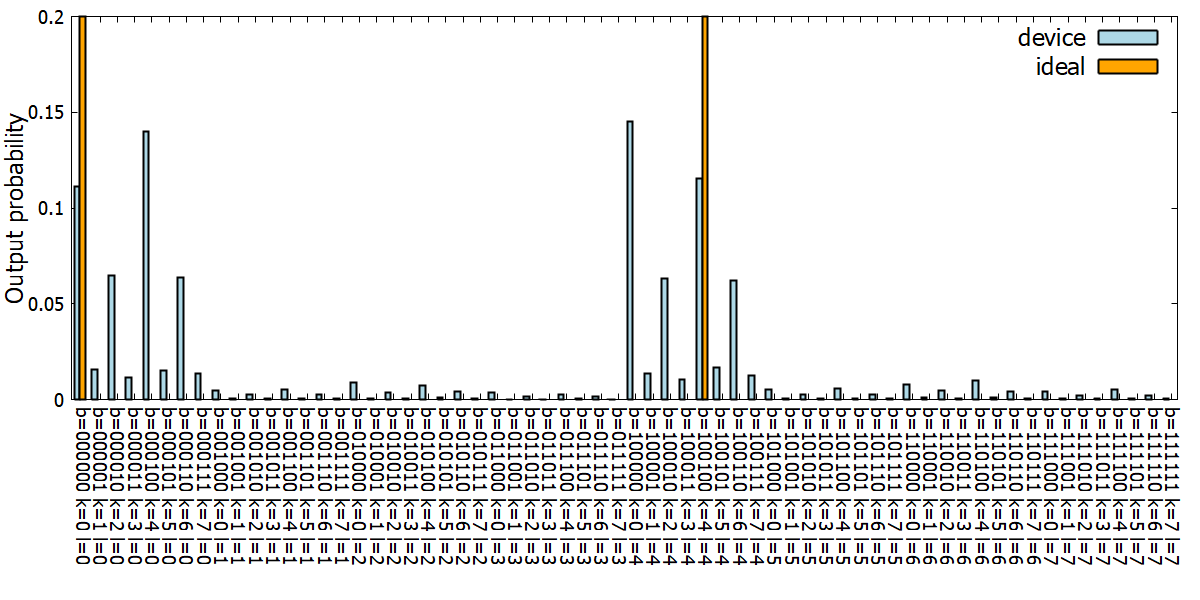} %instance III
		\caption{ Probability distribution of ideal output, IBM Quantum real device output. From the top, the experimenting circuits are I,II and III in Table~\ref{tab:ourinstances}.
		In instances II and III, the values are cut at 0.2 and the ideal probabilities are all 0.5.
	\label{fig:kawasakidist}}
\end{figure}

We compare the output distribution by which we denote $P_{\kawasaki}$ and the ideal distribution $P_{\ideal}$.
Figure~\ref{fig:kawasakidist} shows 
the comparison among the probability distributions of instance I, II and III.

From the graph, clearly the instance I is solved.
In fact, using our post-processing algorithm (Section~\ref{sec:ourpostprocess}) on the instance I,
the success probability is always higher than $99.9\%$ for $K=2,\ldots,20$.
%For instance, the success probability of our post-processing algorithm 
%(Section~\ref{sec:ourpostprocess}) is about 100\% for $K=2,\ldots,20$.
On the other hand, in the instances II and III, although the exact peaks of $P_{\kawasaki}$ and $P_{\ideal}$ are close to each other,
there are some other peaks that are not expected, such as $00100$.
So, in the next section, we discuss the effects of these peaks in post-processing.
Instances IV ,V ,VI have more than $100$ CNOT gates and are not expected to 
output meaningful results.

%\aono{Need KL-divergence?}
%\tanaka{I think not to need}

\subsection{Success Probability Results}

\begin{figure}[htbp]
	\centering
	\includegraphics[width=8cm]{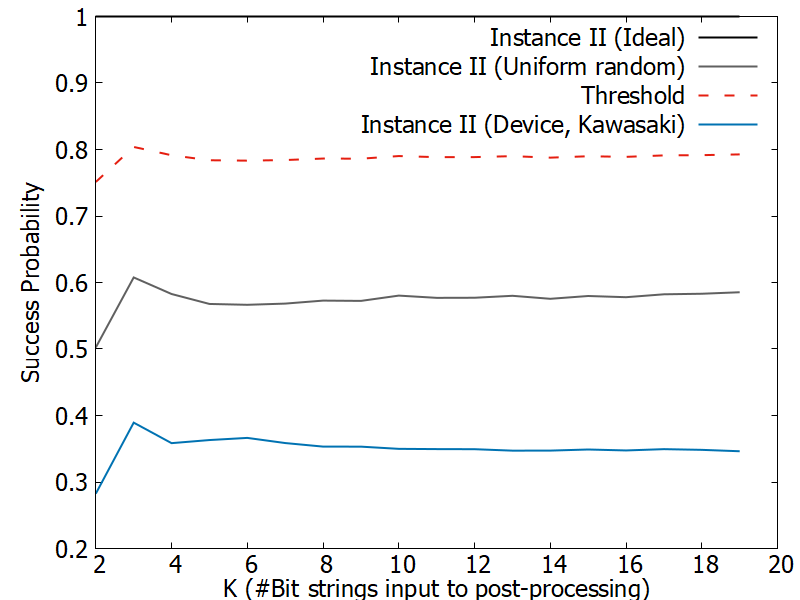} 
	\includegraphics[width=8cm]{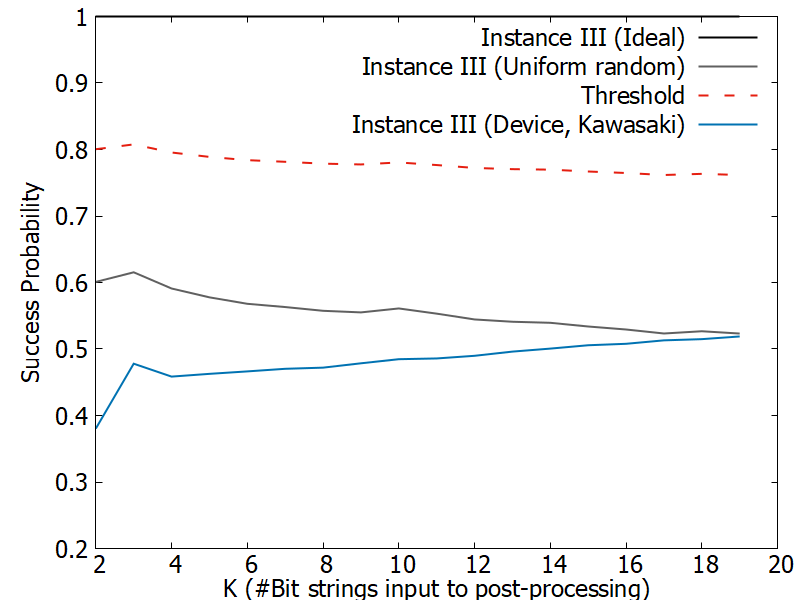} 
		\caption{The experiment success probabilities using our post-processing algorithm (Algorithm~\ref{alg:postprocess}).
		The top and bottom are on instance II and III respectively.
		Note that this result is before using our bit modification algorithm, whose 
		detail and result are explained in Section~\ref{sec:after1bitmodify} 
		and \label{sec:bitselection}.
		Here, $K$ the number of used bit strings sampled from the set of bit strings generated 
		by the IBM Quantum device.
	\label{fig:psucckawasaki}}
\end{figure}

To facilitate a more quantitative comparison, 
we checked the success probability of the post-processing algorithm.
Figure~\ref{fig:psucckawasaki} shows
the success probability of $p_{\ideal}$, $p_{\unif}$ and 
the experimental success probability $p_{\kawasaki}$ (i.e., the probability 
that the return of Algorithm~\ref{alg:postprocess} includes the correct solution) 
of instances II and III, 
using 819,200 samples from the device.
we compared the probabilities without bit string modification in both cases 
as described in Section~\ref{sec:bitselection}.

Figure~\ref{fig:psucckawasaki} depicts a summary of the results.
The success probability from the ideal distribution is almost 1,
and the threshold values are computed by $(1+p_{\unif})/2$.
Unfortunately, the success probabilities of the device outputs are lower than 
that of the uniform.
This can be expressed as follows.
In the instance II and III, the DLP has the solution $z=1$ and 
the vectors that spans the dual lattice are $(0,1)$ and $(1/2,1/2)$.
Thus, $(p,r) = (0,0)$ and $(1/2,1/2)$ from the bit string $\bm{s}_0 = {\tt 00000}$ and $\bm{s}_1 = {\tt 10100}$ in instance II
is the point observed under ideal conditions.
On the other hand, noise bit strings $\bm{s}_2 = {\tt 00100}$ and $\bm{s}_3 = {\tt 10000}$ are frequently observed in experiments.
In particular $\bm{s}_2 $ corresponds to $(p,r)=(1/2,0)$ and the solution $z=0$
that the post-processing algorithm makes mistake.
We think the reason that the error bit sequence $\bm{s}_2$ is higher than the correct solutions ($\bm{s}_0 = {\tt 00000}$ and $\bm{s}_1$) is a concentration of error since 
the Hamming distance relation $d_H(\bm{s}_0,\bm{s}_2) = d_H(\bm{s}_1,\bm{s}_1)=1$.

\subsection{Results after 1bit modification}

\label{sec:after1bitmodify}

To obtain better results, 
we apply our procedure to modify the bits introduced in Section~\ref{sec:bitselection}.
We also apply the modification for the uniform output for a fair comparison.
The probability distribution of the device output before and after modification is supplied in Appendix~\ref{app:deviceprobs}.

Figure~\ref{fig:psucckawasakimodify} shows the comparison of success probabilities.
We can see that the device probabilities have increased, as have the thresholds.
The success probabilities of device outputs are still lower than the 
thresholds.
We emphasize that in instance III, the probabilities of the device become higher than the uniform.
As a result, it appears that if noise levels are reduced, there is a chance of success.
In the following section, we will discuss how much noise we need to reduce through simulation.

\begin{figure}[htbp]
	\centering
	\includegraphics[width=8cm]{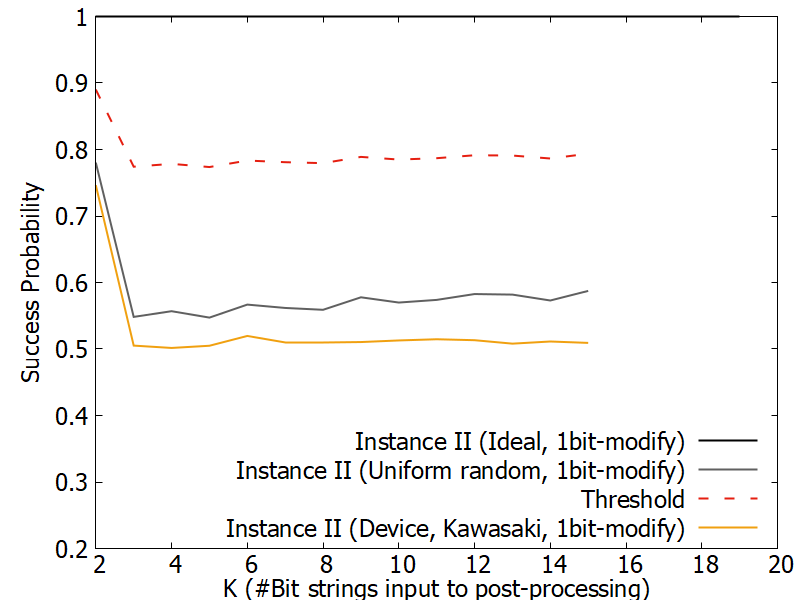} 
	\includegraphics[width=8cm]{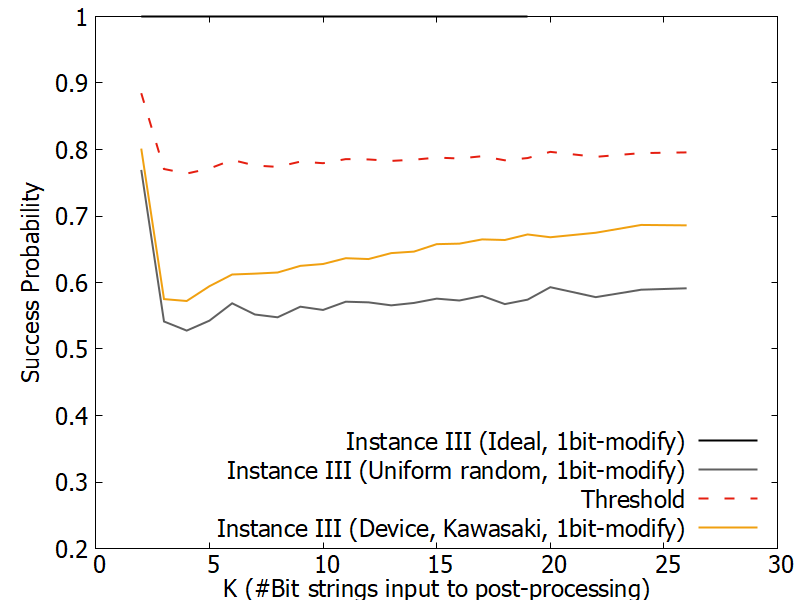} 
		\caption{ The experimented success probabilities of 
		our post-processing algorithm with modified bit string.
	\label{fig:psucckawasakimodify}}
\end{figure}

%% file: expnoisesim.tex
\section{How much noise do we need to reduce}

\label{sec:prediction}

As we can see in the above section, 
the problem level that the latest quantum device can solve is 
between I and II.
This section provides the results of our simulation of noisy devices 
to discuss the near-future of DLP and IFP against quantum devices.

Noise considered in the real quantum devices is represented by many parameters.
To simplify the discussion, we represent the noise by one real number $p_2$ that indicates 
the 2-bit gate depolarizing error.

\begin{figure}[!htbp]
	\centering
\ifthenelse{\boolean{OmitPictures}}{OMITPicture}{
	\includegraphics[width=8cm]{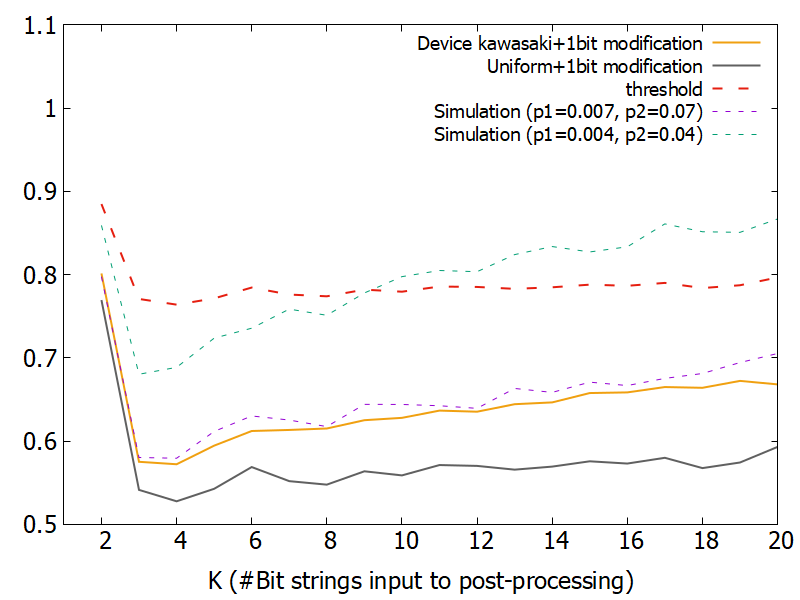} 
	}
		\caption{Comparison of success probabilities.
		The yellow line is the probability of random bits with 1bit modification.
		Because the success probability of ideal output is 1, the threshold (red dot line) is
		computed by $(p_{\unif}+1)/2$.
		The blue line is the probability of real Kawasaki device and it matches to 
		the purple dot line of simulation with $p_2=0.07$.
		To claim success, it is necessary to reduce the noise level of $p_2=0.04$.
	\label{fig:kawasakiandsim}}
\end{figure}

We simulated the circuit of instance III, setting 
1-bit and 2-bit gate depolarizing errors to $0.1\cdot p_2$ and $p_2$, respectively.
Figure~\ref{fig:kawasakiandsim} shows the success probabilities.
Since the success probability of random bits with 1bit modification is about 0.55, the threshold (red dot line) is about 0.8.
On the other hand, the success probability of Kawasaki output with 1bit modification is about 0.6, which does not meet the threshold.
The success probability graph is comparable to that of our simulation with $p_2=0.07$.
It is necessary to achieve $p_2=0.04$ to obtain a success experiment.
We remark that $p_2=0.07$ does not match the claimed CNOT gate error of {\tt ibm\_kawasaki} \cite{IBMQexperience},
because we tried to represent the effect of whole errors in the real device by using a single value.
From the ratio of $p_2$ we think it is necessary to halve the noise
to claim the successful experiments of instance III.

\aono{To be discussed: We also note that a simulation using the noise model published in IBM Q Experience 
outputs a probability distribution very far from the real device output...}

\begin{figure}[htbp]
	\centering
	\ifthenelse{\boolean{OmitPictures}}{OMITPicture}{
	\includegraphics[width=8cm]{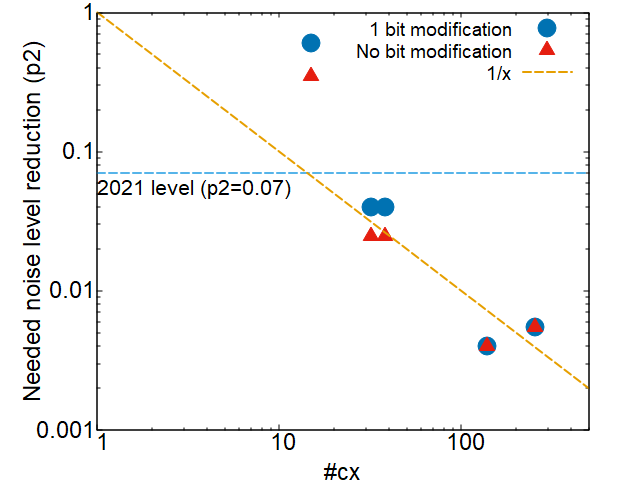} 
	}
		\caption{Comparison between $\# {\tt cx}$ and the necessary noise level $p_2$.
		The blue circles and red triangles are the results with bit string after and before modification, respectively.
		For the points from instances II to V, they all fits on the line of $1/\# {\tt cx}$.
		The horizontal dot line is $p_2=0.07$, the noise level as of 2021.
	\label{fig:cxsizeandnoise}}
\end{figure}

We also simulate the circuits in Table~\ref{tab:ourinstances} 
to find the noise levels we need to achieve to claim success.
The result is summarized in Table~\ref{tab:necessarynoiselevel} and Figure~\ref{fig:cxsizeandnoise}.
$p_{mod}$ and $p_{nomod}$ are maximum of $p_2$ to claim the success of the experiment when 1bit modification is used and unused, respectively.
Here, we decide the success if one of the success probabilities of post-processing for $K=2,\ldots,10$ is higher than the threshold.

The values on the instance II and III look the same, whereas the details are different.
The size of QFT is one of the differences between II and III.
Based on the results of the experiments, we discovered that increasing the QFT size, 
results in increased noise resiliency.
In these cases, it appears that the noise increase caused by the number of gates and the resiliency from the QFT size are balanced.

For the instance IV and V, 
the values $p_{mod}$ and $p_{nomod}$ are completely the same
because the modification algorithm (Algorithm~\ref{alg:modifybits}) have very small effects on the inputs.
Since both instances are from the DLP instance with $p=6$,
the set of legitimate points $S_p$ defined by the equation (\ref{eqn:defsp}) contains 31 points
and the number of corresponding legitimate bit strings are inherently greater than 31.
On the other hand, the possible number of output bit strings is 
$2^{n_x+n_y}=64$ and 256 for instance IV and V, respectively.
Thus, most output are regarded as legitimate such that the true condition is satisfied in Step 2 in Algorithm~\ref{alg:modifybits}.

%Noise lower bound so that the post-processing succeed for at least one of K=2,...,10
% I -> p1=0.06
% II,III -> p1=0.004
% IV -> p1= 0.0004
% V -> p1=0.00055 (large nx and ny implies noisy-resilient?)

%nomod
% I -> p1=0.035
% II,III -> p1=0.0025
% IV -> p1= 0.0004
% V -> p1= 0.00055 (same as modified bits, maybe due to small p)

\begin{table}[!htbp]
\caption{Summary of necessary noise level (as of $p_2$ in simulation)
for the considered instances.
The $\# {\tt cx}$ row is the same as the $\# {\tt cx}\ \textrm{min}$ row in Table~\ref{tab:ourinstances}.
$p_{mod}$ and $p_{nomod}$ are necessary $p_2$ to claim the success experiment 
when 1bit modification is used and unused, respectively.
\label{tab:necessarynoiselevel}}

\begin{center}
	\begin{tabular}{|c|r|l|l|}
		\hline
		        & $\# {\tt cx}$ & $p_{mod}$ & $p_{nomod}$ \\
        \hline
             I & 15 & 0.6 & 0.35 \\
        \hline
            II & 32& 0.04 & 0.025 \\
        \hline
        III & 38 &  0.04 & 0.025 \\
        \hline
        IV & 179 & 0.004 & 0.004 \\
        \hline
        V & 255 & 0.0055 & 0.0055 \\
        \hline
	\end{tabular}
\end{center}
\end{table}

Figure~\ref{fig:cxsizeandnoise} is the log plot of 
$\# {\tt cx}$ and the probability.
We can see the values of instances II to V are on
the line $1/x$.
This explains the weakness of circuits against noise.
Any one of the CNOT gates has an error, the whole computation would be failed.

\noindent
{\bf Future prediction}:~
IBM provides us the history of averaged CNOT gate error rate \cite{IBMQheavyhex} over the past 5 years.
The rate continues to decrease and it falls by roughly half every year.
We can predict when the DLP instances will be solved 
if the total noise level of real quantum devices decreases at the same rate. 
As previously stated, the IBM Quantum ({\tt ibm\_kawasaki}) released in 2021 corresponds to the noise level $p_2=0.07$, which we use to simulate CNOT and single gate errors.
To solve the instances II and III, $p_2=0.04$ is required.
%If the current trend continues, they will be solved by 2022.
If the current trend will continue, they would expect to be solved by a quantum device released in 2022.
Also, in 2025, the noise level will be reached about $p_2=0.07\cdot 2^{-4}=0.004375$ and the instances IV and V are expected to be solved.

%%p2=0.04 - 2022 (II,III)
%%p2=0.02 - 2023 
%%p2=0.01 - 2024 (15A,15B,21A?)
%%p2=0.005 - 2025 IV,V
%%p2=0.0025 - 2026 21B

We also give a prediction on Shor's factoring algorithm based on the number of CNOT gates and a slightly evidenceless assumption 
\begin{equation}
\label{eqn:p2reverse}
p_2 \approx 1/ \# {\tt cx}
\end{equation}
because we have never discussed enough lattice-based post-processing algorithms and bit string modification for the integer factoring situation.

Table~\ref{tab:ourshorinstances} summarized that 
factoring quantum circuits that were considered.
Similar to the DLP circuits, we implement them using {\tt Qiskit}
and compile them using the {\tt transpile} commmand with 100 different seeds.
The columns 
$\#${\tt cx} min and $\#${\tt cx} ave represent the minimum and average of
$\#${\tt cx} of transpiled circuits.

The factoring circuits considered here are textbook proof-of-concept circuits.
That is, they are simplified by using the properties of $a^x\ \mod\ N$.
For example, the second line instance 15B is simplified by using the fact $2^x\ \mod\ 15 \in \{ 1,2,4,8\}$ and $2^4\ \mod\ 15 =1$.
The third line instance 21A is from the circuit by Amico et al.\cite{ASK19}.
The details of the factoring circuits considered are described in Appendix~\ref{app:shorcircuits}.

From the viewpoint of $\# {\tt cx}$, 
15A, 15B and 21A are 2-3 times harder than the DLP instances II and III.
Under our assumption, when $p_2=0.07\cdot 2^{-3}=0.00875$ is achieved in 2024, 15A and 15B could be solved since $1/\# {\tt cx}=0.01$ and $0.0116$, respectively.
The instance 21A has a chance to be solved since $1/\# {\tt cx}=0.00794$.
On the other hand, to solve the larger instance 21B, 
we have to wait for the machine to progress until 2026.

This near term prediction also explains why the existing reports on success experiment 
of integer factoring in this 20 years \cite{VSB+01}-\cite{DLQ+20} except for oversimplified circuits only factor 15 and 21.
It is not enough to execute the proof-of-concept circuits of factoring 15, 
which require a hundred CNOT gates.
The first step will take several years.
We predict larger instances will constantly solved after the first report of complete execution of factoring 15 or 21 including post-processing 
because the growing speed of gate size is polynomial of $\log N$ to the factoring number $N$.
We expect new reports of factoring the numbers greater than 35 will be published frequently after 2025.

\begin{table}[!htbp]
\caption{Summary of quantum circuits for proof-of-concept of Shor's factoring algorithm.
\label{tab:ourshorinstances}}

\begin{center}
	\begin{tabular}{|c|c|c||r|r||r|r|r|}
		\hline
		& Factoring & Bits  & QFT & \multirow{2}{*}{$a$} & \multirow{2}{*}{$\#$Q}& $\#${\tt cx} & $\#${\tt cx}  \\
		& instance & of $N$ & size &&&avg & min \\
		\hline
		\hline
		15A & \multirow{2}{*}{$15$} & \multirow{2}{*}{$4$} & \multirow{2}{*}{$4$} & 7 & \multirow{2}{*}{$9$} & 131.22 & 100 \\
		\cline{1-1} 	\cline{5-5} \cline{7-8} 
		15B & && & 2 &  & 94.42 & 86\\
		\hline
		21A & \multirow{2}{*}{$21$} & \multirow{2}{*}{$5$} & 3 & \multirow{2}{*}{2}  & 8 & 163.99 & 126 \\
		\cline{1-1} 	\cline{4-4} \cline{6-8} 
		21B & && 4 &  & 9 & 448.28 & 365 \\
		\hline
	\end{tabular}
\end{center}

\end{table}

%% file: discussion.tex
\section{Discussion}

\label{sec:discussion}

\rdv{What was the meaning of doing these experiments? What do they tell us about the future of DLP problems on QCs?}

We reported our experiments on the DLP with the IBM Quantum device.
The entire performance of the latest quantum device against DLP including the post-processing algorithm is very limited.
It can solve the smaller 2 bit instance $2^z \equiv 1\ (\mod\ 3)$ (instance I in Table~\ref{tab:ourinstances}) in the sense of median principle on the success probability.
On the other hand, it might reduce the noise by half to solve the larger instance $2^z \equiv 2\ (\mod\ 3)$ (instance II and III in Table~\ref{tab:ourinstances}).

We have predicted the 3 bit DLPs considered in Table~\ref{tab:ourinstances} and factoring in Table~\ref{tab:ourshorinstances} would be solved around 2025.
On the other hand, solving 4 bit or larger instances seem to necessitate the use of a device with quantum error correction.
The importance of noise reduction has been known since the early days of Shor's algorithm~\cite{miquel1996fdq}.
Here, we provide quantitative support by (\ref{eqn:p2reverse}) for that consensus.

\medskip
\noindent
{\bf Threatening RSA-2048}:~
The growing speed of device performance under the assumption from the history of device noises and (\ref{eqn:p2reverse}) is notable.
Assuming the executable number of CNOT gates with negligible errors is doubled every year, it can be expected that a quantum circuit with millions of gates can be executed in the coming decades.
The estimation of Gidney and Eker{\aa} \cite[Table 1]{GE19} claims that about $2.7\times 10^9$ abstract Toffoli gates are required to factor RSA-2048.
Assuming one Toffoli gate is implemented by 5 CNOT gates and several single qubit gates, about $1.35 \times 10^{10} \approx 2^{34} $ CNOT gates
are necessary to implement the circuit.
As a result, it can be expected that a quantum device that threatens integer factoring-based cryptosystems will be developed in the next 30-40 years.

This prediction backs up previous reports predicting that 
the time of compromise RSA-2048 will also arrive in the coming decades.
Sevilla and Riedel \cite{SJ20} make a prediction based on the assumption of exponential progress of physical qubits and gate fidelity, claiming that this is an optimistic scenario.
Their prediction is based on their quantifier of quantum devices that they named generalized logical qubits.
They predicted that a superconducting quantum device capable of solving RSA-2048 (using 4,100 qubits) would be available in the early 2050s, rather than before 2039.
This is more optimistic than expert opinions \cite{PM2019quantum,PM2020quantum} published in 2019 and updated in 2020.
Mosca and Piani say that 90\% of experts predict that there is 50\% or greater chance of a quantum device that can break RSA-2048 in 24 hours being released in the next 20 years.

In addition, our method also shares the difficulty of early prediction of emerging technologies.
That is, the predictions on RSA in the above assumed the exponential growth of some performance values at a glance from historical values.
Due to the small number of datapoints, the evidence supporting our assumptions is not very strong and we need to refine the predictions by taking account into the progress of quantum devices and experiments over the next few years.

We also remark that our prediction method is slightly different from the typical methods in cryptographic research.
To predict threats on RSA and DSA in future, cryptographers typically have used ``bit lengths of 
cryptosystems'', represented by the bit lengths of numbers to be factored in RSA and the modulus in DSA, to measure the security of cryptosystems.
However, this strategy cannot be applied simply to the situation of quantum computing. Despite the advances in technology, the known claimed successful factoring experiments using Shor's circuit on real quantum computers have only been for the values 15 and 21 over a span of 20 years \cite{VSB+01}-\cite{DLQ+20}.
Thus, it is difficult to predict the time of compromise RSA-2048 from only previous experiments.
Therefore, another predicting method was needed, leading us to the present analysis.

\if0
%%Reduce QV part
These predictions are in contrast to the future prediction based on IBM's doubling roadmap \cite{IBMQVdouble} of quantum volume.
The QV is defined by Cross et al. \cite{CBS+19} 
with its approximation $QV \approx 2^{ \min(m,d)}$
such that a machine can solve the heavy output generating problem for a ``square'' circuit of $m = d$ \cite{CBS+19}. Here, $m$ and $d$ represent the width and depth of the quantum circuit under consideration, respectively, implying that QV is based on one side of the smallest square circuit that the machine can handle. 
The challenge is to translate current QV into long-term cryptographic predictions.
Takahashi-Kunihiro \cite{TK06} and H\"{a}ner et al. \cite{HRS17}, for example, proposed an $n$-bit integer factoring circuit using $2n+2$ qubits.
It was predicted that the bit length of factored integers by Shor's circuit would increase by one bit every two years.
Within this prediction, solving RSA-2048 is very unlikely to realize in near-future.

\fi

\medskip\noindent
{\bf Technical future work}:~
Our experiments in this paper employ the standard way to construct the circuits.
To increase the success probability, several techniques can be applied.

In addition to the simplification of modular exponentiation circuits,
the approximation of QFT can also be used \cite{Cop94,NSM20}.
It can reduce the number of gates, i.e., total noise levels, at the expense of accuracy.
Balancing the hardware noise and its inaccuracy is a problem that we hope to address in the near-future.

Quantum error correction (QEC) is under continuing development.
Once implemented, it can also significantly reduce the noise level.
It will be interesting to see how the noise level trend changes as QEC and fault tolerant techniques are applied to actual algorithm execution~\cite{IonQroadmap2020,egan2021fault}.

The post-processing algorithm is also a work in progress, with room for continued improvement.
Based on the set $S_p$ of valid bit strings, we proposed a simple bit modification algorithm.
Other methods should be considered.
Outputs with low levels of noise, in particular, can aid in the recovery of the correct solution in some security-related problems \cite{HS09}.

%% file: missing-lattice.tex
\section{Supporting Technical Materials}

\label{app:missing}

\subsection{Background Theory of our Post-Processing}

\label{app:backtheoryofPP}

This section provides a framework for deriving 
the computational problem (Problem~\ref{prob:duallatticeproblem})
which is implicitly described in several previous works \cite{EH17,Eke16,Eke17}.
We begin by explaining why the noiseless execution of Shor's circuit (Figure~\ref{fig:dlpcircuit}) produces a point close to a lattice point.

We regroup (\ref{eqn:stateafterqft}) by the value of $F = g^x a^{-y}\ \mod\ p$ as 
\[
    \sum_{k,\ell}^{N_x-1,N_y-1} \sum_{F=0}^{p-1} \sum_{\substack{x=0,y=0 \\ g^xa^{-y} \ \mod\ p = F} }^{N_x-1,N_y-1}   e^{2\pi i \cdot \left( \frac{kx}{N_x} + \frac{\ell y}{N_y} \right)  }  \ket{k,\ell} \ket{ F }.
\]
and thus, the probability that we observe $\ket{k,\ell}$ is
\begin{equation}
    \label{eqn:probsumafterqft}
    \sum_{F=0}^{p-1}
    \left| 
\sum_{\substack{x=0,y=0 \\ g^xa^{-y} \ \mod\ p = F} }^{N_x-1,N_y-1}   e^{2\pi i \cdot \left( \frac{kx}{N_x} + \frac{\ell y}{N_y} \right)  } \right|^2
\end{equation}
up to the normalization constant.
To our best knowledge, there is no closed simple formula 
to compute or approximate this. 

However, we can expect for a pair $(k,\ell)$,
if there exists a constant $c_{k,\ell,F}$ that does not depend on $x$ and $y$,
for $\forall (x,y)$ satisfying $g^xa^{-y} \ \mod\ p = F$, it satisfies
\[
    \frac{kx}{N_x} + \frac{\ell y}{N_y} - c_{k,\ell,F}
    \ \mod\ 1 \approx 0.
\]
Then the probability factor $|\cdot|^2$ in (\ref{eqn:probsumafterqft}) is high for the pair.
Here, $\mod\ 1$ transforms a real number to a number within the range $(-0.5,0.5]$.
In other words, the above $(k,\ell)$ satisfies 
For any pairs $(x,y),(x',y')$ 
satisfying $g^xa^{-y} \ \mod\ p = g^{x'}a^{-y'} \ \mod\ p$,
\[
    \begin{array}{l}
    \displaystyle
    \left( \frac{kx}{N_x} + \frac{\ell y}{N_y} \right)
    -\left( \frac{kx'}{N_x} + \frac{\ell y'}{N_y} \right)
    \ \mod\ 1\\
    \displaystyle
    =
    \left( \frac{k}{N_x} , \frac{\ell}{N_y} \right)
    \cdot (x-x',y-y') \ \mod\ 1 \approx 0.
    \end{array}
\]

This condition can be interpreted using the terminology of lattices.
Since the function 
$g^xa^{-y} \ \mod\ p$ have two periods $(z,1)$ and $(p,0)$,
the above $\left( \frac{k}{N_x} , \frac{\ell}{N_y} \right)$
satisfies 
\[
    \left( \frac{k}{N_x} , \frac{\ell}{N_y} \right)
    \cdot (x,y) \approx \mathbb{Z} 
\]
for all the lattice point $(x,y) \in L \cap [0,N_x-1]\times [0,N_y-1]$ where the lattice $L(B)$ is spanned by $(z,1)$ and $(p,0)$.
Thus, we can expect $\left( \frac{k}{N_x} , \frac{\ell}{N_y} \right)
$ is an approximation of a point in $L^\times \cap [0,1)^2$
if the observed pair has no quantum errors.

We explain the derivation of an instance $(B,\bm{y},\rho)$ of bounded distance decoding (BDD) problem 
and how the solution implies the desired DLP solution $z$.
The matrix representation of $B$ and its dual $D$ are given as 
\[
B =	
\left[
\begin{array}{cc}
	p-1 & 0 \\
	z & 1 \\
\end{array}
\right]
\ \ \ \textrm{and} \ \ \ 
D = \frac{1}{p-1}
\left[
\begin{array}{cc}
	1 & -z \\
	0 & p-1 \\
\end{array} 
\right].
\]

Suppose we carry out $K$ shots on a quantum device 
and have bit-strings $\bm{s}_1,\ldots,\bm{s}_K$.
Also, we let the $i$-th point interpreted from the $i$-th vector
as
\begin{equation}
	\tag{\ref{eqn:measurep}}
	\bm{p}_i :=
	(p_{i},r_{i}) = \left( \frac{k_i}{ N_x} ,\frac{\ell_i}{N_y} \right)
\end{equation}

There exists a short error vector 
$(\varepsilon_{i},\eta_{i}) $ so that 
$(p_i+\varepsilon_i ,r_i+\eta_i) \in L^\times$.
Decomposing into coordinates, 
for each measurement, there exists integers $a_i,b_i$ and they satisfy
\[
p_{i} = \frac{a_i}{p-1} - \varepsilon_{i}\ \ 
\textrm{and}\ \ 
r_{i} =  \frac{-a_i z }{p-1} +b_i +  \eta_{i}
\]
and therefore, erasing $a_i$ we obtain 
the fundamental relation 
\begin{equation}
    \label{eqn:dlpfundamentalindual}
    z\cdot p_{i} + r_{i} -b_i = - z\cdot \varepsilon_{i} +
    \eta_{i}
\end{equation}

Revealing $z$ from given sequence of $p_{i}$ and $r_{i}$
is the computational problem (Problem~\ref{prob:duallatticeproblem}) 
that we have to solve.
One straightforward way is a procedure 
that calls a BDD oracle.
Consider the BDD instance given by the $(K+1) \times K$ matrix 
\begin{equation}
	\tag{\ref{eqn:defB}}
    \begin{small}
    B :=
    \left[
    \begin{array}{c}
        \bm{b}_1 \\
        \bm{b}_2 \\
        \bm{b}_3 \\
            \vdots \\
        \bm{b}_{K+1}
    \end{array}
    \right] 
    =
    \left[
    \begin{array}{ccccc}
    	p_{1} & p_{2} & \cdots & p_{K} \\
    	1 & 0 & \cdots & 0 \\ 
    	0 & 1 & \cdots & 0 \\
    	\vdots & \vdots & \ddots & \vdots \\ 
    	0 & 0 & \cdots & 1 
    \end{array}
    \right],
    \end{small}
\end{equation}
the target vector 
\begin{equation}
    \tag{\ref{eqn:defy}}
    \yvec = (r_{1},r_{2},\ldots,r_{K}),
\end{equation}
and the radius $\rho$ to be optimized.
To find suitable $\rho$, a refined determinant analysis is necessary, 
which we will describe in Appendix~\ref{app:latticeanalysis}.

We can see that there exists $\xvec \in L$ 
close to $\bm{y}$ satisfying
\begin{equation}
	\label{eqn:closevector}
\yvec - \xvec = \left(  -z \varepsilon_{1} + \eta_{1}, -z\varepsilon_{2} + \eta_{2},\ldots,
-z\varepsilon_{K} + \eta_{K}   \right).
\end{equation}

We explain how to recover $z$.
We first remark that finding the combination coefficient $a_i$ of 
the lattice vector $\bm{x} = \sum_{i=1}^{K+1} a_i \bvec_i $, $z$ 
is easily recovered since $z=-a_1$.
Our implementation via lattice algorithms is slightly technical as follows.

The first operation is to transform the lattice basis (\ref{eqn:defB}) 
to its full-rank form since a BDD subroutine typically takes a full-rank matrix basis as its input.
For instance, there is an efficient algorithm (for example, see \cite[Sect.~6]{Bre11} or \cite[Sect.~2.6.4]{Coh13}) that outputs a pair of a unimodular matrix $U$ and a square basis matrix $B'$ so that satisfies
\[
    \begin{array}{ll}
    B' := 
    \left[
    \begin{array}{c}
        \bm{b}'_1 \\
        \bm{b}'_2 \\
        \bm{b}'_3 \\
            \vdots \\
        \bm{b}'_{K}
    \end{array}
    \right] \\
    = 
    UB
    =
    \left[
    \begin{array}{cccc}
    	u_{1,1} & u_{1,2} & \cdots & u_{1,K+1} \\
    	\vdots & \vdots & \ddots & \vdots \\ 
    	u_{K,1} & u_{K,2} & \cdots & u_{K,K+1} 
    \end{array}
    \right]
    \left[
    \begin{array}{c}
        \bm{b}_1 \\
        \bm{b}_2 \\
        \bm{b}_3 \\
            \vdots \\
        \bm{b}_{K+1}
    \end{array}
    \right].
    \end{array}
\]
In particular, each $\bm{b}'_j$ is expressed as $\sum_{i=1}^{K+1} u_{j,i} \bm{b}_i$.

Then, a BDD subroutine with input $(B',\bm{y},\rho)$ returns a set of lattice vectors.
Let $\bm{x} = \sum_{j=1}^{K} c_j \bvec'_j $ be a vector in the set.
Here, a typical BDD subroutine returns its combination coefficients $c_j$.
If not, we can easily compute them by simple matrix multiplication.
By the relation 
\[
    \sum_{j=1}^{K} c_j \bvec'_j = \sum_{j=1}^{K} c_j \sum_{i=1}^{K+1} u_{j,i} \bm{b}_i,
\]
the coefficient of $\bm{b}_1$ is $\sum_{j=1}^K c_j u_{j,1}$ 
and it is $-z$, the negative of desired DLP solution candidate.

\subsection{Refined Determinant Analysis}

\label{app:latticeanalysis}

This section gives our refined theoretical analysis of 
the covolume of the lattice defined by (\ref{eqn:defB}),
which is used in our BDD implementation in relatively small dimensions.
The technique is generally called the point-counting method
and we apply it to the dual lattice of $B$.

\begin{theorem}
    \label{thm:covolB}
	Let $B$ be the matrix of the form (\ref{eqn:defB}).
	In particular, assume that 
	all $p_{i}$ are fractions of the form 
	$a_i / 2^{m_i}$ where $a_i$ is odd integer or zero.
	For $p_{i}=0$, we set $a_i=m_i=0$.
	Then, the covolume of the lattice spanned by the rows of $B$ is
	\[
		\covol(B) = 2^{-m_j}
	\]
	where $m_j$ is the maximum of all $m_i$'s.
\end{theorem}

\mybeginproof 
Let $\bm{w} = (w_1,\ldots,w_K)$ be a vector in the dual lattice
and consider the condition the vector must satisfy.
By the definition of dual lattice, $\langle \bm{w},\bm{b}_i \rangle \in \mathbb{Z}$ for all $i=1,\ldots, K+1$.
Thus, $\bm{w}$ must be an integer vector by the condition for $i=2,\ldots,K+1$.
Consider the condition for $i=1$, it must be 
\[
\sum_{i=1, i\ne j}^K w_i \cdot \frac{a_i}{2^{m_i}} 
+ w_j \cdot \frac{a_j}{2^{m_j}} \in \mathbb{Z}.
\]
After fixing all $i$ except for $j$, 
there exists just one $w_j$ in $2^{m_j}$ integers,
so that the inner-product is an integer.
Thus, we have $\vol(D) = 2^{m_j}$ and $\vol(B) = \vol(D)^{-1} = 2^{-m_j}$.
\myendproof

In our situation $\max m_i = n_x$ (QFT size) holds with high probability since $p_i = a_i / 2^{n_x}$ are considered as random numbers from the observations.

\subsection{Selecting Radii in BDD}

\label{app:selectingradii}

Computing the lattice covolume, 
one can try to set the searching radius $\rho$ in the BDD subroutine
so that the computational time of post-processing is feasible.
The number of lattice points in the ball $\ball_K(\bm{y},\rho)$ should be small,
for this purpose, i.e., bounded by some constant or sub-polynomial function of the lattice dimension.
As a result, determining a relationship between the number of lattice points and the radius is crucial.
In typical situations, the Gaussian heuristic plays an important role.
However, we discover that the heuristic does not hold in our DLP situation,
posing a new unsolved problem.

The Gaussian heuristic says there exists approximately 
$\vol(\ball_K)/\covol(L)$ lattice points in the intersection $\ball \cap L$
for a $K$-dimensional full-rank lattice.
In our situation, the number can be computed as 
\[
	\frac{\vol( \ball_K(\bm{y},\rho) )}{\covol(L)}
	= 2^{n_x} \cdot \frac{\pi^{K/2}\cdot \rho^K}{\Gamma(K/2+1)}.
\]
Thus, setting the radius $\rho_K' = N^{1/K}\cdot  \rho_K$ where 
\begin{equation}
    \label{eqn:ghradius}
    \rho_K = 2^{-n_x / K} \Gamma(K/2+1)^{1/K} \pi^{-\frac{1}{2}},
\end{equation}
the expected number of lattice points is about $N$.

\begin{figure}[htbp]
	\centering
\ifthenelse{\boolean{OmitPictures}}{OMITPicture}{
	\includegraphics[width=7cm]{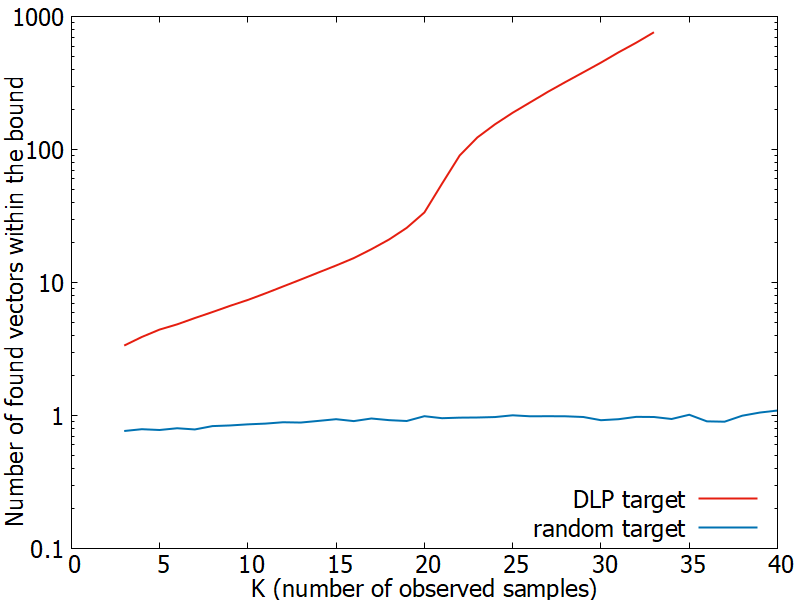}
}
	\caption{The averaged numbers of vectors within the ball $\ball_K(\bm{y},\rho)$ 
	where $\bm{y}$ is set from the DLP (red) and set randomly (blue).
	The radius $\rho$ is set by  (\ref{eqn:ghradius}) so that 
	the average number is expected to be 1.
	\label{fig:numcompare}
	}
\end{figure}

Eker\r{a} \cite[Sect.~5.2.1]{Eke19}'s estimation compares 
the above radius and an asymptotic upper bound $\Delta_K$ to $\| \bm{y} - \bm{x} \|$ in the $K$-dimensional space.
Then, estimate how many samples $K$ is required so that the lattice-based postprocessing can be completed in a reasonable amount of time. 
However, based on our preliminary experiments, we discovered that the strategy needs to be modified.
We simulated noiseless quantum computation 
and sample many bit-strings for the DLP instance V.
We generate 10,000 sets of BDD instances 
defined by (\ref{eqn:defB}) and (\ref{eqn:defy}), for each dimension $K$
and count the number of lattice points within $\ball_K(\bm{y},\rho)$
where $\rho$ is defined by (\ref{eqn:ghradius}), corresponding to $N=1$.
For the control, we also count the number of lattice points when the target point $\bm{y}$ is randomly generated from $[0,1)^K$.
The result is depicted in Figure~\ref{fig:numcompare}.
We can see the number in the ball grows exponentially in the DLP case
while the random case keeps about 1.

Hence, we think we cannot use the Gaussian heuristic simply 
in the post-processing of Shor's algorithm for DLP.
Either lattices or target vectors are not random.
The reason can be explained as follows.
Suppose there exists a lattice vector $\bm{x}$ in the ball.
Then, the neighborhood lattice vectors $\bm{x} \pm \bm{e}_i \pm \bm{e}_j \pm \cdots$
are very likely in the ball where $\bm{e}_i=(0,\ldots,0,1,0,\ldots,0)$ is the unit vector.
As a result, we believe that using the Gaussian heuristic to set the BDD radius is not a good strategy in this situation.
How to set the searching radius or searching space for large $K$ 
should be an open problem.
This is why, in the main section, We propose using post-processing with small $K$.

We remark an another setting method by using the covering radius could be considered,
but may not be very useful.
The covering radius of a lattice $L$ is defined by the minimum radius $c$ of the ball so that 
the set $\cap_{\bm{x} \in L} \ball_K(\bm{x},c) = \textrm{span}(L)$.
In other words, the BDD instance $(B,\bm{y},c)$ has always a solution for any $\bm{y} \in \textrm{span}(L)$.
Thus, setting $\rho_K$ by some upper bound of the covering radius, the return of BDD subroutine is always not empty.
However, there could find an exponential number of lattice points even for $K=2$.
Let us consider the following extreme situation
with $K=2$, $p_{1,1}=N_x^{-1}$ and $p_{2,k}=0$.
The covering radius of the lattice is $c = \frac{1}{2}\sqrt{1 + N_x^{-2}}$
which is achieved by the target vector $\bm{y} = (0,0.5)$.
On the other hand, for the target vector $\bm{y} = (0,0)$ 
the ball of radius $c$ contains an exponential number of lattice points.
Thus, we decided to keep using the simple radius from the Gaussian heuristic
and use an extra CVP subroutine if the BDD subroutine returns the empty set.

\if0
\subsection{Miscellaneous: On the Selection of $K$}

\label{sec:selectingK}

In Section~\ref{sec:ourpostprocess},
we claimed that using large $K$ should not be a very good strategy due to the exponential time of postprocessing algorithm (supporting experiments are given in Appendix~\ref{sec:appendixradii}).

\begin{figure}[htbp]
    %From aono's 20210723data folder
	\centering
\ifthenelse{\boolean{OmitPictures}}{OMITPicture}{
	\includegraphics[width=8cm]{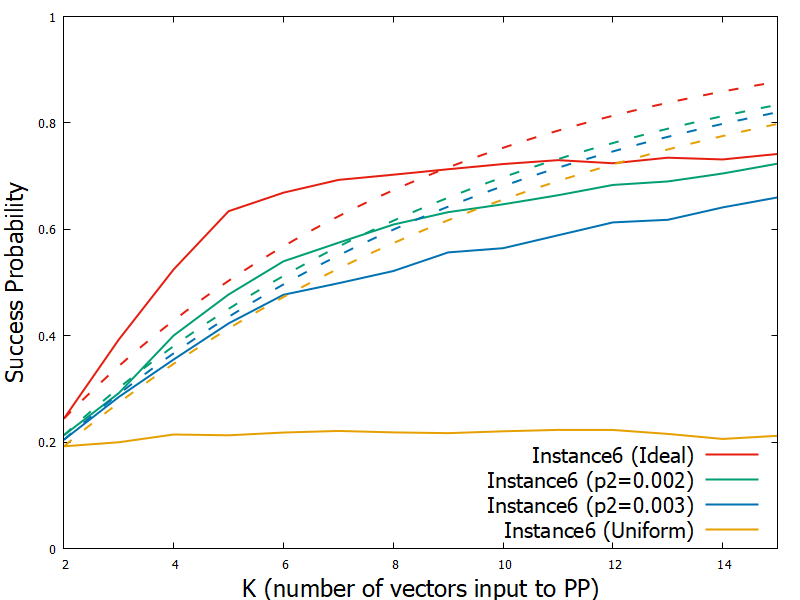}
}
	\caption{Comparison among success Probabilities of post-processing algorithm using $K$ samples (bold lines), and $K/2$ trials of the algorithm using 2 samples. \label{fig:psucc6}}
\end{figure}

We provide another evidence to use small number of samples by the success probability of post processing.
Figure~\ref{fig:psucc6} shows the success probability from the simulated bit-strings.
The bold lines are success probabilities of post-processing algorithms using $K$ samples.
Here, we only remove the zero vectors, but 
do not modify the bit-strings as in Section~\ref{sec:bitselection}.

On the other hand, the dot lines are 
the total success probability of $K/2$ trials 
of post-processing algorithms using 2 vectors.
In other words, for each device success probability $p_{\device,2}$ the graphs show $1-(1-p_{\device,2})^{K/2}$.

From the graph, we conclude that using many samples 
as possible as we can is not a very good strategy.

\fi

\subsection{Device Probabilities after 1 Bit Modification}

\label{app:deviceprobs}.

This section gives supporting materials of Section~\ref{sec:after1bitmodify}.
We give a detailed example of our modification algorithm 
and probability distribution results 
of the {\tt ibm\_kawasaki} device and the distribution after the 1bit modification on instances II and III.

Since the both instances II and III consider the DLP instance of $p=2$,
the set of a legitimate set is
\[
    S_2 = \{ (0,0), (0.5,0), (0.5,0.5)\}.
\]
The corresponding bit-strings are {\tt 00000}, {\tt 00100}, and {\tt 10100} in instance II,
and {\tt 000000}, {\tt 000100} and {\tt 100100} in instance III, respectively.
We should explain the bit-string $b_0b_1b_2b_3b_4$ from the Qiskit output is interpreted 
as $(k,\ell)=(4b_2+2b_3+b_4,2b_0+b_1)$ in instance II.
Also, the bit-string $b_0b_1b_2b_3b_4b_5$ in instance III is interpreted as 
$(k,\ell)=(4b_3+2b_4+b_5,4b_0+2b_1+b_2)$.

Table~\ref{tab:modifybitsinstance2} shows 
the summary of 1 bit modification for an instance with $p=3$.
For instance II, its ideal distribution takes $(0,0)$ and $(0.5,0.5)$ with probabilities 0.5,
which correspond to the bit-strings {\tt 00000} and {\tt 10100}.
The device frequently outputs bit-strings that are modified to incorrect bit-string {\tt 00100}
and we can see the probabilities of 
{\tt 10100} (correct) and {\tt 00100} are very close to each other.
This is one reason why the experiment failed.

\begin{table}[!htbp]
\caption{Summary of 1 bit modification (Algorithm~\ref{alg:modifybits}) on instance II.
\label{tab:modifybitsinstance2}}

\begin{center}
	\begin{tabular}{|c|l|}
        \hline
        Input & Output Candidates \\
        \hline
        {\tt 00000} & {\tt 00000} \\
        {\tt 10000} & {\tt 00000}, {\tt 10100} \\
        {\tt 01000} & {\tt 00000} \\
        {\tt 00100} & {\tt 00100} \\
        {\tt 00010} & {\tt 00000} \\
        {\tt 00001} & {\tt 00000} \\
        \hline
        {\tt 10100} & {\tt 10100} \\
        {\tt 01100} & {\tt 00100} \\
        {\tt 00110} & {\tt 00100} \\
        {\tt 00101} & {\tt 00100} \\
        \hline
        {\tt 11100} & {\tt 10100} \\
        {\tt 10110} & {\tt 10100} \\
        {\tt 10101} & {\tt 10100} \\
        \hline
        Others & $\perp$ \\
        \hline
	\end{tabular}
\end{center}
\end{table}

\begin{figure}[htbp]
\ifthenelse{\boolean{OmitPictures}}{OMITPicture}{
	\centering
	\includegraphics[width=8cm]{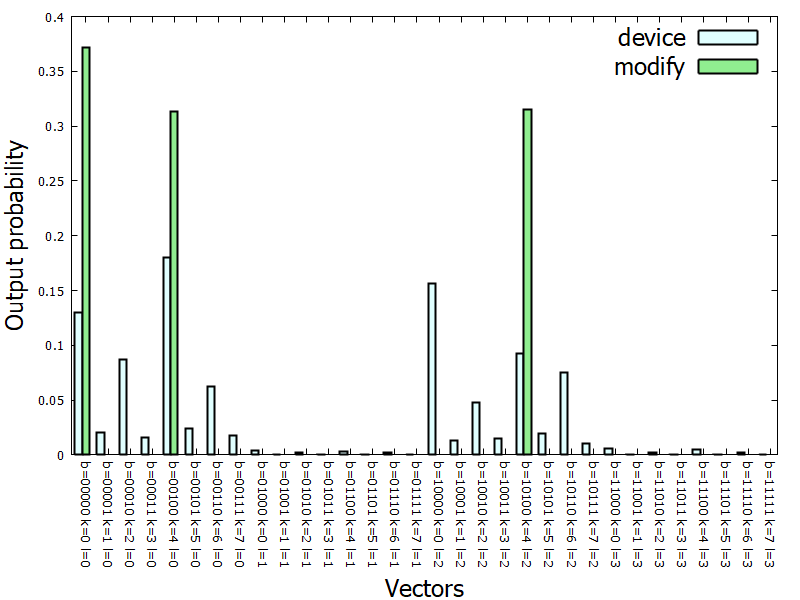} 
	\includegraphics[width=8cm]{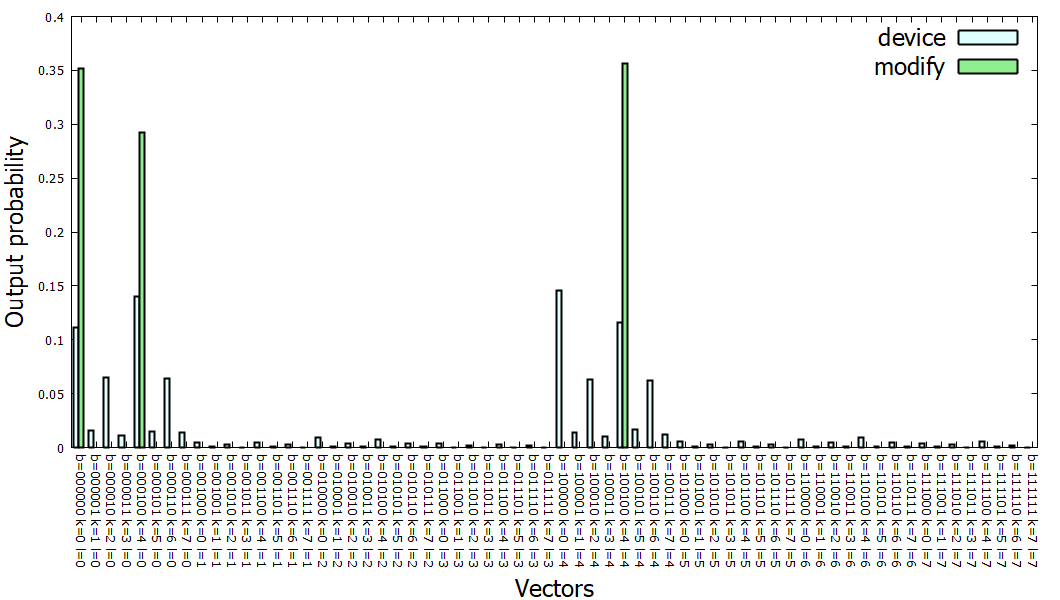} 
	}
		\caption{ 
	\label{fig:devicebitmodify}
	    Comparison between the probability distribution of the {\tt ibm\_kawasaki} outputs
	    and after 1 bit modification.}
\end{figure}

Figure~\ref{fig:devicebitmodify} shows 
the probability distribution before and after modification 
of the {\tt ibm\_kawasaki} device outputs.
After the modification, we can see the instance III has better results because the probability of the correct bit-string ({\tt 100100}) is higher than the other ({\tt 000100}),
and this explains the difference in success probabilities (Figure~\ref{fig:psucckawasaki} and \ref{fig:psucckawasakimodify}).

%% file: appenddlpcircuits.tex
\subsection{Construction of our DLP circuits}

\label{app:dlpcircuit}

This section provides the detailed construction of the
modular exponentiation part of the DLP circuit 
experimented in Section~\ref{sec:experimentquantum}.

%\begin{widetext}
\begin{figure*}[htbp]
	\centering
\ifthenelse{\boolean{OmitPictures}}{OMITPicture}{
	\includegraphics[width=16cm]{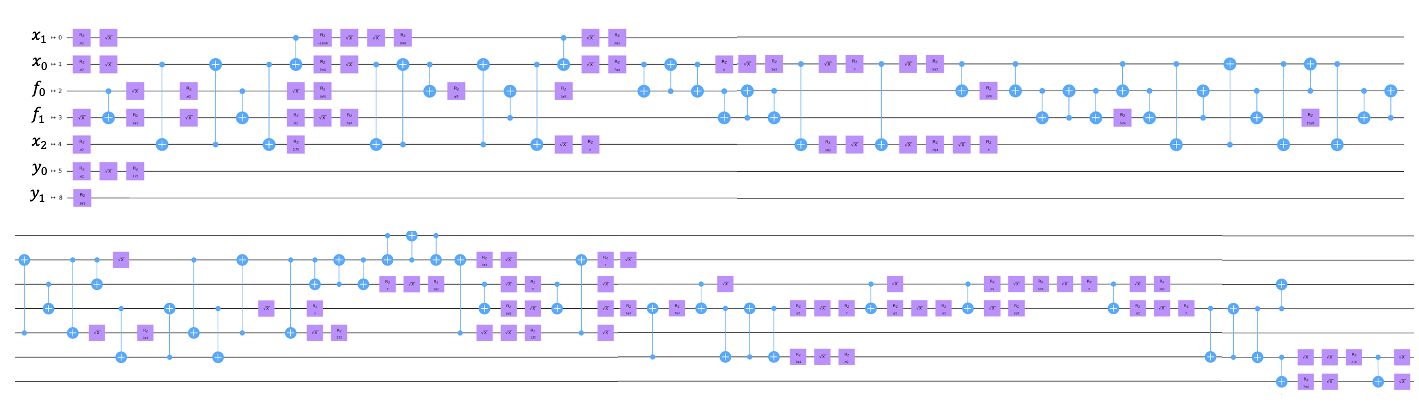}
}
	\vspace{5mm}

\ifthenelse{\boolean{OmitPictures}}{OMITPicture}{
	\includegraphics[width=16cm]{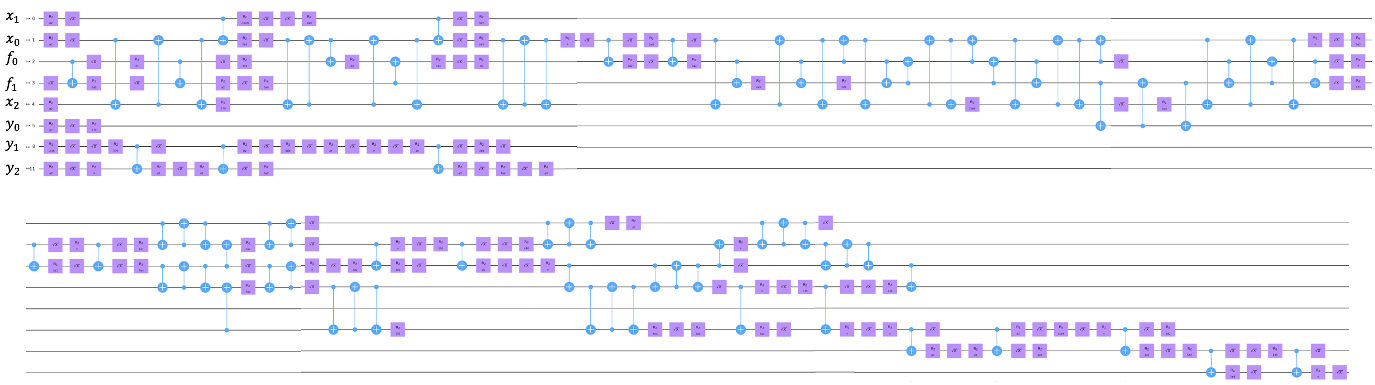} 
}
\caption{Our transpiled circuits of instance II (Top) and III (Bottom) to solve the DLP instance $2^z \equiv 2\ (\mod\ 3)$.	The last observing gates are omitted.}
	\label{fig:dlp23circuit}
\end{figure*}
%\end{widetext}

%% file: appendshor.tex
\subsection{Construction of our factoring circuits}

\label{app:shorcircuits}

This section provides the detailed construction of the
modular exponentiation part of Shor's factoring circuit
considered in Table~\ref{tab:ourshorinstances}
the experimental reproducibility.

We first remark that 
the four circuits that we considered are proof-of-concept versions.
That is, they are oversimplified by techniques that are not scalable.

\begin{figure}[htbp]
	\centering
\ifthenelse{\boolean{OmitPictures}}{OMITPicture}{
	\includegraphics[width=6cm]{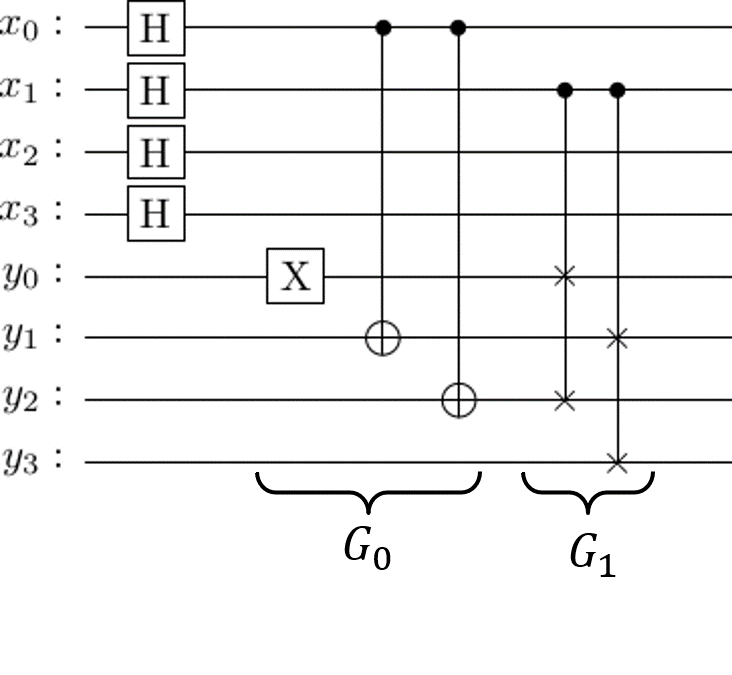} 
}
		\caption{ 
	\label{fig:factoring15A}
	Modular multiplication gadgets
	for a proof-of-concept circuit of Shor's factoring algorithm with parameters $(a,N)=(7,15)$ and size of QFT is 4.
	}
\end{figure}

\medskip\noindent
{\bf Construction of 15A}~
Factoring circuit with $(a,N)=(7,15)$ is traditionally discussed in Vandersypen et al. \cite{VSB+01} and recently, Monz et al. \cite{MNM+16} and Duan et al. \cite{DLQ+20} reported experiments with using semi-classical QFT.
Our circuit (Figure~\ref{fig:factoring15A}) follows \cite{VSB+01}.
It computes $Y=(y_32^3+y_22^2+y_12^1+y_02^0) = 
7^{x_32^3+x_22^2+x_12^1+x_02^0}\ \mod\ 15$.
The gadget $G_0$ computes $2^{x_0}$,
and $G_1$ computes $\times 4^{x_1}\ \mod\ 15$ since $7^2 = 4\ \mod\ 15$.
The computations on $x_2$ and $x_3$ can be omitted since $7^4\equiv7^8\equiv 1\ (\mod\ 15)$.

\begin{figure}[htbp]
	\centering
\ifthenelse{\boolean{OmitPictures}}{OMITPicture}{
	\includegraphics[width=6cm]{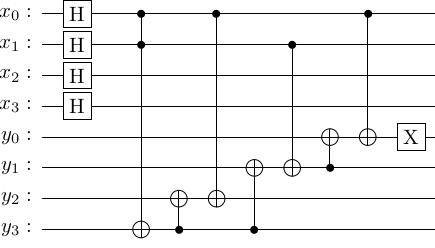} 
}
\caption{ 
	\label{fig:factoring15B}
	Modular multiplication gadgets
	for a proof-of-concept circuit of Shor's factoring algorithm with parameters $(a,N)=(2,15)$ and size of QFT is 4.
	The number of CNOT gates after transpiling is slightly reduced to the circuit of Figure~\ref{fig:factoring15A}.
	}
\end{figure}

\medskip\noindent
{\bf Construction of 15B}~
For the parameters $(a,N)=(2,15)$, the size of factoring circuit can be slightly reduced by using the information that $2^x\ \mod\ 15 \in \{ 1,2,4,8 \}$.
The logical relation between $Y=(y_32^3+y_22^2+y_12^1+y_02^0)$
and $X=x_1 2+x_0$ are written as
\[
    \begin{array}{ll}
        y_3 = x_0 \cdot x_1, & y_2 = x_0\cdot \overline{x_1} \\
        y_1 = \overline{x_0} \cdot x_1 & y_0 = \overline{x_0}\cdot \overline{x_1}
    \end{array}
\]
and they are interpreted to the following formulas
and the circuit in Figure~\ref{fig:factoring15B}.
\[
	\begin{array}{l}
	y_3 = x_0 \cdot x_1  \\
	y_2 = y_3 \oplus x_0   \\
	y_1 = y_3 \oplus x_1   \\
	y_0 = \overline{y_1 \oplus x_0}  \\
	\end{array}
\]

\begin{figure}[htbp]
	\centering
\ifthenelse{\boolean{OmitPictures}}{OMITPicture}{
	\includegraphics[width=6cm]{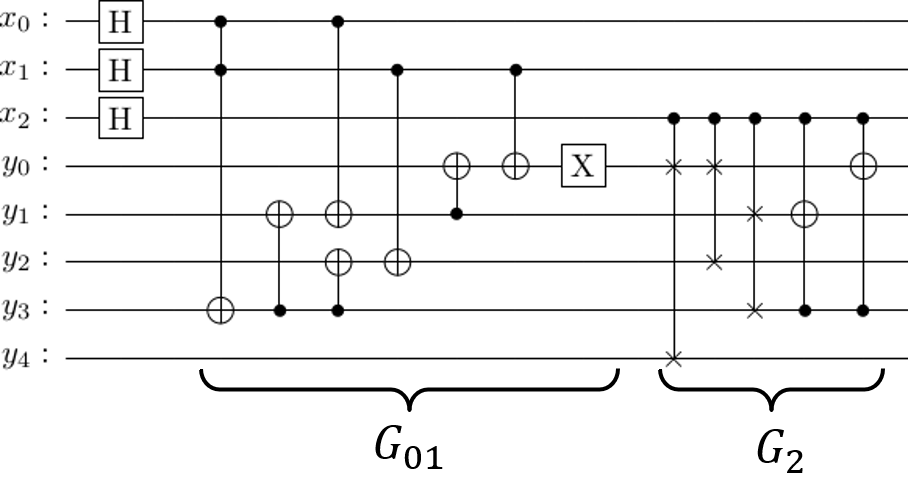} 
}
\caption{ 
	Modular multiplication gadgets
	for a proof-of-concept circuit of Shor's factoring algorithm with parameters $(a,N)=(2,21)$ and size of QFT is 3.	
	\label{fig:factoring21A}}
\end{figure}

\medskip\noindent
{\bf Construction of 21A}~
We construct a factoring circuit with parameters $(a,N)=(2,21)$ and size of QFT 3.
We follow the construction of Amico et al.\cite{ASK19}.
Our circuit to compute 
$Y=(y_42^4 + y_32^3+y_22^2+y_12^1+y_02^0) = 
2^{x_22^2+x_12^1+x_02^0}\ \mod\ 21$ 
is displayed in Figure~\ref{fig:factoring21A}.
In the gadget circuit $G_{01}$, it computes 
$Y=2^{x_12^1+x_02^0}$ by using the precomputed information that $2^{X}$ for $X=0,1,2,3$ is $1,2,4,8$ respectively.
This gadget is the same as circuit 15B.
Then, the next gadget computes $\times 2^{4x_2}\ \mod\ 21$.
The trick here uses the fact for $Y=\{ 1,4,8\}$, the result of 
$16Y\ \mod\ 21$ and $16Y\ \mod\ 63$ are the same.
This allows us to use the bit-rotation technique described in Section~\ref{sec:simplegadgets}, which is implemented by the first three cswap gates in $G_2$.
After the cswap gates applied, $Y={\tt 01000}=8$ corresponding to $X={\tt 101}=5$ is only the value to modify.
The last two ccx gates modify it to $2^5\ \mod\ 21=11$.

\begin{figure}[htbp]
	\centering
\ifthenelse{\boolean{OmitPictures}}{OMITPicture}{
	\includegraphics[width=6cm]{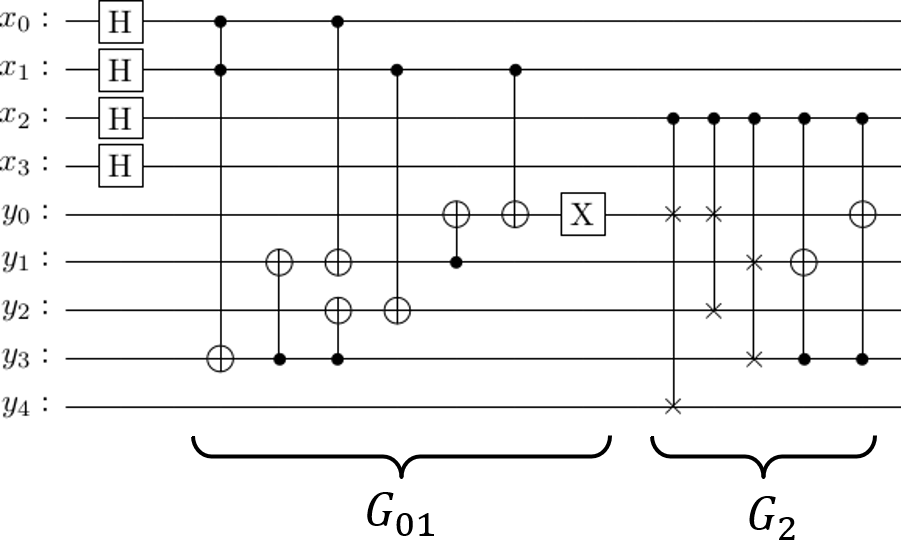} 
	\includegraphics[width=6cm]{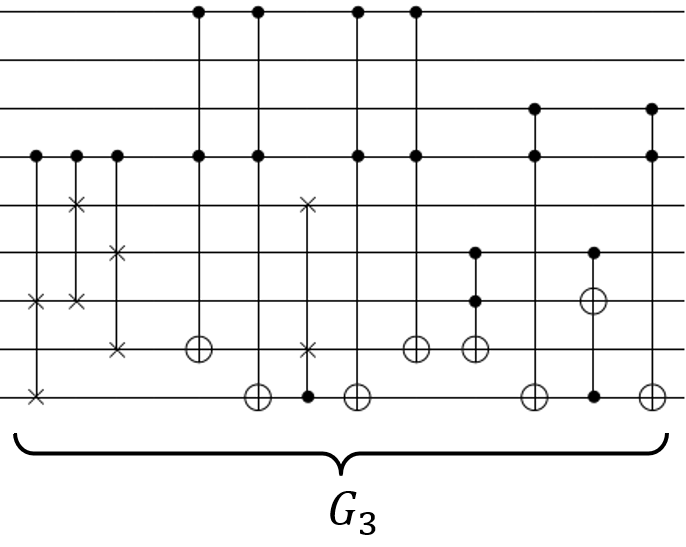} 
}
		\caption{ 
	Modular multiplication gadgets
	for a proof-of-concept circuit of Shor's factoring algorithm with parameters $(a,N)=(2,21)$ and size of QFT is 4.	
	\label{fig:factoring21B}}
\end{figure}

\medskip\noindent
{\bf Construction of 21B}~
A circuit with parameters $(a,N)=(2,21)$ and size of QFT 4 is constructed by a similar method to that of 21A.
In Figure~\ref{fig:factoring21B}, the gadgets $G_{01}$ and $G_2$ are the same as Figure~\ref{fig:factoring21A}.
To construct the $(\times 2^{8x_3}\ \mod\ 21) = (\times 4^{x_3}\ \mod\ 21)$ circuits, we use the equivalence between 
$4Y\ \mod\ 21$ and $4Y\ \mod\ 63$ for $Y = 1,2,4,16$.
The first three cswaps change the values of $Y$
as described in Table~\ref{tab:needtomodifyvalue}.
The remained part transforms $Y=2,14$ to $11,2$ respectively
without changing other values.

\begin{table}[!htbp]
\caption{Values that are necessary to be modified after 
the first three cswaps in $G_3$.
\label{tab:needtomodifyvalue}}

\begin{center}
	\begin{tabular}{|r|r|r|}
        \hline
        Input $Y$  & $Y$ after cswaps & correct value \\
        after $G_2$ & in $G_3$ & $4Y\ \mod\ 21$ \\
        \hline
        1 & 4 & 4\\
        \hline
        2 & 8 & 8 \\
        \hline
        4 & 16 & 16 \\
        \hline
        8 & 2 & 11 \\
        \hline
        11 & 14 & 2\\
        \hline
        16 & 1 & 1 \\
        \hline
	\end{tabular}
\end{center}
\end{table}

